\documentclass{aa}
\usepackage{graphicx}
\usepackage{txfonts}
\usepackage{aalongtable}

\begin{document}

%\headnote{Research Note}

\def\SBS{SBS~1520+530 }
\def\hyp{Hyperz }
\def\ho{H$_0$ }
\def\hirac{HiRAC }
\def\sex{SExtractor }
\def\kms{km\,s$^{-1}$}

\title{The DPOSS II compact group survey: first spectroscopically confirmed candidates\thanks{Based on observations obtained at the NTT ESO telescope on La Silla (Chile)}}

\subtitle{}

\author{E. Pompei\inst{1}\and R.R. de Carvalho\inst{2} \and A. Iovino \inst{3}}

\offprints{Emanuela Pompei: epompei@eso.org}

\institute{
          European Southern Observatory,
Alonso de Cordova 3107, Vitacura, Casilla 19001, Santiago 19, Chile
	\and
INPE/MCT,
Av. dos Astronautas 1758,
S. J. Campos, SP 12227-010, Brazil;
reinaldo@das.inpe.br 
         \and
Osservatorio Astronomico di Brera
Via Brera 28, I-20121 Milano, Italy;
iovino@brera.mi.astro.it}

\date{Received: May 15, 2005/ Accepted:}

\abstract{This paper presents the results of a pilot redshift survey of 18 candidate compact groups from the distant DPOSS survey that extends the available surveys of compact groups of galaxies to $z \sim 0.2$
in redshift, mainly Hickson Compact Groups and Southern Compact Groups.}
{The goal of our survey was to confirm group
membership via redshift information and to measure the characteristic
parameters of a representative, albeit small, sample of DPOSS survey
groups.}
{Of the 18 candidates observed, seven are found to be indeed isolated
compact groups, i.e. groups with 3 or more concordant members and with
no neighbouring known cluster, while 7 are chance
projection configurations on the sky. Three remaining candidates,
despite having 3 or more concordant member galaxies, are located in
the neighbourhood of known clusters, while another candidate turned out to be a
dense sub-condensation within Abell 0952.}
{The median redshift of our 7 confirmed groups is $z \sim 0.12$, to be
compared with a median redshift of 0.03 for
the local sample of compact groups by Hickson.  The typical group size is
$\sim$ 50 kpc, and the median radial velocity
dispersion is 167 \kms, while typical crossing times range from
0.005 H$_{0}^{-1}$ to 0.03 H$_{0}^{-1}$ with a median value of 0.018
H$_{0}^{-1}$, all similar to the values usually found in the
literature for such structures in the local universe. The average
mass-to-light ratio for our groups, M/L$_{B}$, is 92{\it h},
higher than the value found for nearby Hickson compact groups
but lower than that found for loose groups.  Our results suggest that, 
once full redshift information for its members
becomes available, the DPOSS sample will provide a reference sample to study the properties of compact groups beyond the local universe.
}{}
\keywords{galaxies:clusters:general -- method:spectroscopy -- compact groups --}

\maketitle

\titlerunning{DPOSS II compact groups survey}

%
%________________________________________________________________
\section{Introduction}

Compact groups (hereafter CGs) are well known systems, ever since the discovery by Edouard
Stephan of the first one in 1877, at a time when we did not even know
about the expansion of the universe and about the existence of other galaxies
outside our own. These galactic systems drew attention due to their
small angular scale; their sizes are comparable to the mean distance
between their member galaxies. Physically, we naively classify these
systems as having low mass, high projected density, and low velocity
dispersion. Galaxy-galaxy interactions (e.g. close tidal
encounters) and mergers are therefore likely to dominate their
evolution. Until now, however, signs of interactions in CGs selected
from the Hickson catalogue, the most widely studied catalogue of such
objects (Hickson 1982), were found to be uncommon and traces of
mergers to be rare (Zepf 1993). When present, these features are
mainly found in spiral-dominated groups.\newline 
Subsequent studies by Mendes de Oliveira
et al (1994) showed that, while interactions (i.e. encounters
which do not disrupt the galaxies) between galaxies
in compact groups are quite common, mergers remain rare, to the level of 6$\%$ of
the group galaxies. Addtional studies by Proctor et al. (2004) showed that
the stellar population of the early-type member galaxies of nearby, z $\le$ 0.03, 
compact groups is quite old (Proctor et al, 2004). 
This led to doubt the existence of compact groups as physically bound system,
but detection of diffuse intra-group matter in 75$\%$ of HCGs (Ponman et al., 1996)
confirmed that these are indeed bound, self-gravitating
structures.  The question then arises about the origin and evolution of compact groups:
how can they survive for so long?\newline
Studies of the environment of HCGs
(de Carvalho et al. 1997; Ribeiro et al. 1998) have shown that compact
groups can be divided into three categories, namely: real compact
groups, systems composed of a core+halo structure, and sparse groups.
Objects belonging to different categories have different surface
density profiles, and we observe a propensity for higher activity
level in lower velocity dispersion groups. These differences are also
reflected in their X-ray emission: groups detected in X-ray have higher galaxy density
than groups without detectable X-ray emission. Moreover early type
galaxies are more centrally concentrated in X-ray emitting groups
than in the non-emitting ones; and finally groups dominated by late type spirals do 
not show X-ray emission (Zabludoff \& Mulchaey, 1998).

The different properties of the group categories identified
by de Carvalho et al. (1994, 1997) might be interpreted as an evolutionary
scheme in which the groups form in a looser concentration of galaxies,
followed by a period of strong evolution with merging episodes. They
then settle into a more quiescent phase and finally end up as
isolated field ellipticals (Coziol et al., 2004).
However, recent studies by Proctor et al. (2004) and Mendes
de Oliveira et al. (2005) find that early type galaxies
in compact groups are older than field galaxies of the same type and similar to
cluster galaxies. This means that formation of early type galaxies
in groups by galaxy-galaxy
mergers must have happened a long time ago (of the order of a few
Gyrs). On the other hand, compact groups have a very short crossing time, 
of the order of a few percent of the Hubble time, making it unclear how
groups dominated by early type galaxies are observable today, because
they should have disappeared a long time ago. A different possibility
might be that compact groups are quite young configurations and that
we are observing different stages of an ongoing merging
process at the same time. If this is true, then
we might wonder what a search for compact groups at increasing redshift
would produce: would we find an higher number of interacting groups than
at present time? Would we find an increase in the activity of the member galaxies?
Would we find changes in the group's physical characteristics, like velocity
dispersion, mass, radius, and crossing time?

To achieve these goals requires a complete sample of compact groups whose
observational and statistical biases are well understood. 
A new sample of 459 compact group candidates with a median expected
redshift of $\sim$ 0.12, i.e. 10 times higher than the typical
crossing time of present-day compact groups, has been selected with an automatic algorithm
applied to the Digitized Second Palomar Observatory Sky Survey II (herafter DPOSS II) 
galaxy catalogues (Iovino et al. 2003 and de
Carvalho et al.  2005). We refer the reader to these cited papers
for detailed information on the sample, however, we report here the
selection criteria for the sake of completeness:
\begin{itemize} 
\item{} {\it richness}: n$_{member}$ $\ge$ 4 in the magnitude interval
$\Delta$mag$_{comp}$ = m$_{faintest}$-m$_{brightest}$, with the constraint
$\Delta$mag$_{comp}$ $\le$ 2. Here n$_{member}$ is the number of member galaxies
and m$_{brightest}$, m$_{faintest}$ are the magnitude of the brighest and the faintest
group members respectively.
\item{} {\it isolation}: R$_{isol}$ $\ge$ 3R$_{gr}$, where R$_{isol}$ is the distance
from the centre of the smallest circle encompassing all group galaxies to the
nearest non-member galaxy within 0.5 magnitudes of the faintest group member.
R$_{gr}$ is the radius of the smallest circle emcompassing all the galaxies of
the group.
This criterion avoids finding small aggregates of galaxies within a larger
structure, e.g. a cluster.
\item{} {\it compactness}: $\mu_{gr}$ $<$ $\mu_{limit}$, where $\mu_{gr}$ is the mean
surface brightness within the circle of radius R$_{gr}$ and $\mu_{limit}$ = 24 mag arcsec$^{-2}$
in r band. For comparison, Hickson uses a $\mu_{limit}$ = 26 mag arcsec$^{-2}$, which would
have increased the contamination of the sample to 80$\%$, against the current estimate
of 27$\%$.
\end{itemize}
The magnitude difference criterion is considerably stricter
than Hickson's ($\Delta$mag$_{comp}$ $\le$ 3), meaning that we have a lower
contamination rate, 10$\%$ against 27$\%$, but also reduced completeness.\newline

A first step in exploiting such a sample is to
define via spectroscopic follow-up how many groups are indeed bound,
selecting subsamples of groups with three/four galaxies sharing the
same recession velocity.  In addition, the spectroscopic data allow us
to estimate the dynamic characteristics of the groups and assess the
level of activity in their galaxy population. Here we present the
first results from a pilot study carried out at the 3.58m New Technology
Telescope (NTT) at La Silla Observatory. In the next section we 
describe the observation and data reduction, in Sect. 3 we
present our results and in Sect. 4 our findings.
\newline

Throughout the paper, a $\Lambda$CDM cosmology ($\Omega_{M}$=0.3;
$\Omega_{\Lambda}$=0.7) and H$_{0}$=67 \kms Mpc$^{-1}$ have
been used.\newline

\section{The data}

\subsection{Observations and data reduction}

The targets were selected from the published DPOSS II compact group
sample by Iovino et al. (2003) on the basis of available 
observational windows. They represent a fair sample of the total catalogue
published in Iovino et al.(2003); in fact a {\it k-s} test on the observed sample 
and the complete DPOSS II
group catalogue shows that the two populations are the same with a probability
of 96.1$\%$. Notice also that the Iovino et al. (2003)
sample is a subset (with a few exceptions) of the larger sample of
Carvalho et al. (2005), which presents the final complete sample of DPOSS
candidate compact groups.  These group galaxies were observed with the
NTT telescope and the ESO Multi Mode Instrument (hereafter EMMI) in
spectroscopic mode in the red arm, equipped with
grism $\#$2 and a slit of 1.5$\arcsec$, under clear/thin cirrus
conditions and grey time. The MIT/LL new red detector, a mosaic
of 2 CCDs 2048 x 4096,  was binned by 2
in both spatial and spectral direction, with a resulting dispersion of
3.56$\AA$/pix, a spatial scale of 0.33$\arcsec$/pix, an instrumental
resolution of 322 \kms, and a wavelength coverage from 3800 to
11000 $\AA$. When possible, two or more galaxies were placed together
in the slit, whose position angle was constrained by the location
of galaxies in the sky and almost never coincided with the
parallactic angle. Exposure times varied from 720s to 1200s per spectrum, and
two spectra/target were taken for each galaxy to ensure reliable
cosmic ray subtraction.
The full log of the observations is given in Table 1.

\begin{table*}
\caption{Observing log}
\label{Obs}
\begin{center}
\begin{tabular}{llllll} \hline\hline
Group name       & Exp. time                    & $<$seeing$>$ & FLI$^{a}$ & $<$X$>^{b}$  & Sky transparency$^{c}$ \\ 
                 & (seconds)                    &              &           &           & \\ \hline
PCG100355+190454 & 2 x 900 (A,D); 2 x 720 (B,C) & 0.9$\arcsec$ & 100      & 1.5        & CLR/THN\\
PCG101345+194541 & 2 x 900 (A,B,C,D)            & 0.8$\arcsec$ & 30       & 1.6        & CLR/THN\\
PCG103349+225324 & 2 x 900 (A,D); 2 x 720 (B,C) & 0.9$\arcsec$ & 100      & 1.9        & CLR/THN\\ 
PCG103959+274947 & 2 x 720 (A,B); 2 x 900 (C,D) & 1.0$\arcsec$ & 80       & 1.9        & CLR \\
PCG104338+281711 & 2 x 720 (A,B); 2 x 900 (C,D) & 1.2$\arcsec$ & 40       & 1.7        & THN \\
PCG104538+175826 & 2 x 720 (A,B); 2 x 900 (C,D) & 1.0$\arcsec$ & 90       & 1.5        & CLR/THN\\
PCG110941+203320 & 2 x 900 (A,B,C,D)            & 1.2$\arcsec$ & 90       & 1.6        & CLR/THN\\     
PCG114233+140738 & 2 x 900 (A,B,C,D)            & 0.9$\arcsec$ & 20       & 1.5        & CLR/THN\\     
PCG114333+215356 & 2 x 900 (A,B,C,D)            & 1.0$\arcsec$ & 100      & 1.6        & CLR/THN\\
PCG121157+134421 & 2 x 900 (A,B,C,D)            & 1.2$\arcsec$ & 60       & 1.6        & THN \\    
PCG121252+223519 & 2 x 900 (A,B,C,D)            & 0.8$\arcsec$ & 100      & 1.7        & CLR/THN\\
PCG121516+153357 & 2 x 720 (A,B,C,D)            & 1.0$\arcsec$ & 30       & 1.4        & CLR/THN\\         
PCG130157+191511 & 1 x 720 (A,B,C,D)            & 0.9$\arcsec$ & 70       & 1.5        & CLR \\  
PCG130926+155358 & 2 x 900 (A,B,C,D)            & 0.7$\arcsec$ & 80       & 1.8        & CLR \\    
PCG145239+275905 & 2 x 720 (A,B); 2 x 1200 (C,D)& 0.7$\arcsec$ & 90       & 1.9        & CLR \\
PCG154930+275637 & 2 x 900 (A,B,C,D)            & 1.2$\arcsec$ & 60       & 1.9        & THN \\    
PCG161754+275834 & 2 x 900 (A,B,C,D)            & 0.7$\arcsec$ & 90       & 1.8        & CLR \\
PCG170458+281834 & 2 x 900 (A,B,C,D)            & 1.0$\arcsec$ & 50       & 1.9        & CLR/THN\\ \hline
\multicolumn{6}{l}{$^a$ \small FLI = fraction of illuminated moon} \\
\multicolumn{6}{l}{$^b$ \small average airmass of the group} \\
\multicolumn{6}{l}{$^c$ \small CLR = clear; THN = thin cirrus}
\end{tabular}
\end{center}
\end{table*} 
 
On-site data reduction was performed using the EMMI spectroscopic
quick look tool available at La Silla Observatory (Pompei et al.,
2004) and refined later using the MIDAS data reduction package\footnote{Munich
Image Data Analysis System, which is developed and maintained by the
European Southern Observatory}. Our
steps include bias subtraction, flat-field correction, wavelength
calibration, cosmic ray filtering, sky subtraction, and correction by
atmospheric extinction, but no flux calibration, as our nights were not
all of photometric quality. Flat field correction was good up to
2$\%$, except in the reddest part of the spectra (redward of
8000$\AA$), where fringing becomes significant (up to 4$\%$ from 
peak-to-peak variation). As a consequence, absorption lines redward of
8000$\AA$ were not considered for the redshift measurement.  A 
two-dimensional dispersion solution was obtained using the arc frames
taken in the afternoon. No arc was taken during the night, since EMMI
flexures are less than 1 pixel over the full 360 degree instrument
rotation.
A third order polynomial was used for the dispersion direction, while
a second order one was employed to correct for the geometrical
distortion along the spatial direction. An upper limit of 0.16$\AA$
was found for the rms of the wavelength solution.\newline 
Sky subtraction was performed on the two-dimensional rectified
spectrum of each target and the r.m.s in the background of the sky
subtracted spectra varies from 5 to 10$\%$, which is the worst for nights
with a full moon.  The atmospheric absorption feature at $\sim$ 7600
$\AA$ was not corrected and any line falling close or in it was not
considered for the redshift measurement.  The two spectra available
for each galaxy
were collapsed perpendicular to the dispersion direction, in order to measure
the FWHM of the galaxy, on average 4.3$\arcsec$. To obtain the final
one-dimensional, wavelength calibrated galaxy spectrum an extraction
window of 3 x FWHM was always used. An exception to this rule is
represented by two galaxies very close to each other: in this case the
biggest non-overlapping window has been chosen, which was never
smaller than 4.3$\arcsec$.\newline 
The two one-dimensional spectra available for each galaxy were averaged
together at the end of the reduction, giving an average S/N of $\sim$
30 (grey time) or $\sim$ 10-15 (almost full moon) per resolution
element at 6000$\AA$.\newline

In some cases, nearby galaxies happened to fall in the slit together
with the candidate member galaxies, so their spectra were reduced and
extracted following the same procedure used for the target galaxies.
Radial velocity standards from the Andersen et al. (1985) paper were
observed with the same instrumental set-up used for the target
galaxies; in addition to this, we also used galaxy templates with
known spectral characteristics and heliocentric velocity available
from the literature, i.e.  M32, NGC 7507, and NGC 4111.

\subsection{Redshift measurements}

The IRAF\footnote{IRAF is distributed by NOAO, which is operated by AURA,
Inc, under cooperative agreement with the NSF} packages {\it xcsao} and {\it emsao} have been used to
measure the galaxy redshifts by means of cross-correlation method
(Tonry \& Davis, 1979), where good results were obtained with galaxy spectra
dominated by emission lines or by absorption lines.  For spectra
dominated by absorption lines, we used galaxy templates and stellar
radial velocity standards, while for emission line dominated spectra we
used a synthetic template generated by the IRAF package {\it
linespec}.  Starting from a list of the stronger emission lines
(H$\beta$, [OIII], [OI], H$\alpha$, [NII], [SII]), the package creates
a synthetic spectrum, which is then convolved with the instrumental
resolution.\newline
Following the discussion by Kurtz \& Mink (1998), we
performed visual checks of the complete galaxy sample, in order to
understand the reliability limit of the automatic redshift
determination. This is given by a {\it confidence parameter}, {\it r},
defined for the first time in Tonry \& Davis (1979), which is basically the
ratio between the height of the true peak of the correlation function
and the average peak of a spurious function (called {\it remainder function}
in Tonry \& Davis). We found that all redshifts with a confidence parameter {\it r}
$\ge$ 5 are reliable, but measurements with 2.5 $\le$ {\it r} $\le$ 5 need
to be checked by hand.  Measurements with {\it r} $\le$ 2.5 are not
reliable. All the confirmed member galaxies in our sample had
a {\it r} $>$ 3.5.

In some cases {\it emsao} failed to correctly identify the emission
lines, which happened each time some emission lines were contaminated
by underlying absorption.  When this happened, we measured the
redshift by gaussian fitting of the strongest emission lines visible
and took the average of the results obtained from each line.
If two or more lines were blended, the IRAF command {\it deblend}
within the {\it splot} package was used.\newline

The error quoted for the recession velocity is the quadrature sum of two
terms: the first is the error in the dispersion solution and the
second is the error in the velocity estimate, defined as the scatter
in the measurement resulting from the use of different templates, or as
the scatter given by the gaussian fitting of different emission lines.
The recession velocity errors varied between 15 and 100 \kms, depending 
also on the S/N of the target spectrum.\newline

Correction to heliocentric recession velocity values was obtained with
the IRAF task {\it rvcorrect} in the package {\it noao.rv}.\newline

After checking the literature, we found no overlapping data in the 2dF and
6dF survey, and only one of our galaxies has a published redshift in
Nasa Extragalactic Database (hereafter NED), PCG114333+215356B. The quoted 
redshift is z = 0.130 (Tovmassian
et al., 1999), while the one we measured is z = 0.1319$\pm$0.0002, i.e. $\sim$ 600
\kms difference. On the other hand the
published value is quoted there without any errors, and addressing the
reader to a forthcoming paper on the optical spectroscopy, which at the
time of writing this article, was not yet available. Another comparison can be done
with the paper of Miller et al. (2002), where they measure redshifts
of galaxies in Abell 0952. Unfortunately, only one galaxy is common to
both samples, PGC101345+194541A, at $\alpha$ = (10:13:45.11) and
$\delta$ = (+19:45:44.4). The two velocity measurements, 33923$\pm$40 \kms
in our case and 33981$\pm$52 \kms for Miller, agree within
the experimental errors.\newline 

Table 2, which is available only in electronic form, lists the galaxy name (column 1), J2000 coordinates from the
DPOSS II survey (columns 2 and 3), heliocentric radial velocity and
velocity error (columns 4), type of velocity measurement (from
absorption lines ``abs'', from emission lines ``em'', from a
combination of both ``mix''; column 5), and the emission lines detected in
the spectrum, if any (column 6).

\section{Results}

In this section we present the results obtained in our pilot survey.
We first discuss group membership and the spectral
properties of the member galaxies in detail . We then discuss
statistical properties of our sample more generally: velocity
dispersion, crossing time, {\it M/L} ratio and typical size.  While it is difficult
to produce statistically significant results with only 18 groups, it
is nevertheless interesting to examine the trends.
We also searched the environment of each group, checking for the
possible presence of a nearby known cluster, with the goal of ascertaining
whether the group is really isolated or if it belongs to a larger
structure.

Finally we note that sometimes other nearby galaxies, close on
the sky to our target galaxy but not included as member galaxies in
the original group definition, entered in our slit and happened to
have a concordant velocity with other members of the group. In order
to avoid a contamination of the group selection function or of the
group properties, we decided in the following to calculate all the
quantities taking into account only the candidate members, as shown on
the finding charts on the paper from Iovino et al. (2003).\newline

\subsection{{\it Group membership}}

Following the paper by Hickson et al. (1992), we decided to consider
as {\it bona fide} confirmed groups all the candidates with at least
three galaxies, whose relative velocity difference was $\Delta$v $\le$
$\pm$1000 \kms from the median velocity of the group.\newline

The opportunity of such a choice is confirmed by the plot in Fig.
1, which shows the distribution around the median velocity of the
group for all galaxies (both concordant and discordant) and originally
classified as group members. All the spectroscopically confirmed
member galaxies are within 1000 \kms of the median velocity of the
group while the discordant ones are off the scale.  This result is
very similar to the one found by Hickson for the HCGs catalog.

\begin{figure}[h]
\includegraphics[angle=-90,width=9cm]{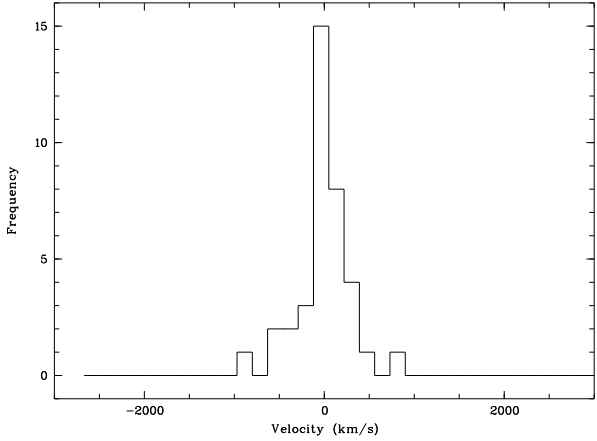}
\caption{Distribution of the difference in velocity from the median
velocity of the group for all the galaxies cataloged to belong to
the same group.}
\label{Fig. 1}
\end{figure}

To calculate the median group velocity and radial velocity dispersion,
we used the biweight estimators of location and scale (Beers et al.,
1990), which takes all relevant cosmological effects into account.
Of a total of 18 candidate compact groups, we found 11 concordant
and 7 discordant objects, for a total of 60$\%$ success rate.
The radial velocities of our groups range from cz = 23800 to 44586
\kms, with a median cz $\sim$ 37534 \kms. These values should be
compared to the corresponding ones for the HCGs: radial velocities
range from 1380 to 41731 \kms, with a median cz $\sim$ 8889 \kms.

With the redshift information of each group in hand, we first
explored the group environment using NED. We decided to take all the available
cluster catalogues into account, with the exception of the Zwicky one, as most of
its clusters lack redshift measurements. In our analysis, we also
included the DPOSS II cluster catalogue (Gal et al., 2003), which is
more homogeneous and covers the same depth and area of our group
survey, and whose clusters have a reliable photometric redshift
estimate. \newline

To find neighboring clusters, we adopted a search radius of
15$\arcmin$, i.e. $\sim$ 3$\arcmin$ bigger than the Abell radius of a
cluster placed at the distance of the our furthest group, z=0.148.
Once we have a list of possible cluster neighbours, we checked if any of
them has a measured redshift. If so, we further refined the
search using a redshift criterion by assuming that a group is close to
a cluster if the difference in redshift between the two is $\Delta$z
$<$ 0.01.  This corresponds to a velocity difference of 3000 \kms, one
order of magnitude above the typical dispersion of compact groups and $\sim$ 2.5
times wider than the biggest velocity dispersion measured for clusters
($\sim$ 1200 \kms; Zabludoff et al., 1993).\newline

Accordingly our candidate groups are then divided in three classes: A,
B, and C.  Candidate groups belonging to class A are confirmed and isolated systems;
candidate groups belonging to class B are confirmed but close on the sky to
larger structures to which they might be associated. Candidate groups
belonging to class C are all the targets with less than 3 concordant
galaxies, and thus not groups according to our definition.
Of the 11 confirmed groups, we found seven that we consider Class A,
while four are close to a cluster. One candidate turned out to be a
dense sub-condensation within Abell 0952, while the remaining three
are located in the outskirts of known clusters. 
In Table 3 we show the group name (column 1), the coordinates of the centre of the group
(columns 2, 3), the number of concordant galaxies (column 4), the mean
redshift of the group (column 5), its class (column 6) and any
neighbour that has been found (column 7).  No cluster name means none
has been found within the search radii used.
It is interesting to note that there is only one candidate group with
4 discordant members, PCG121157+134421.

\setcounter{table}{2}
\begin{table*}
\caption{\label{dposs_groups} Classification of the DPOSS II compact groups and neighbouring
large scale structures. n represents all galaxies which fulfilled the
velocity criteria, irrespective of the original catalogue of group galaxies, while the number quoted
in parentheses is the number of concordant galaxies that were originally part of the
DPOSS catalog. When a (2+2) is given for n (members), it means that the candidate group is composed of two concordant pairs of galaxies.}

\begin{tabular}{llllllll}
\hline\hline
Group name        &  $\alpha$ & $\delta$   &  n        & $<$z$>$   & Group class & Cluster?   & Notes  \\ 
                  & (J2000.0) & (J2000.0)  &  members  &           &             &            &        \\ \hline
PCG100355+190454  & 10 03 55  &  +19 04 54 &  4        & 0.1076$\pm$0.0002 &  A          &  -         &                                      \\ 
PCG101345+194541  & 10 13 45  &  +19 45 41 &  9(4)     & 0.1121$\pm$0.0002 &  B          & Abell 0952 & group in the centre of the cluster   \\
PCG103349+225324  & 10 33 49  &  +22 53 24 &  2+2      &   -       &  C          &  -         &                                      \\
PCG103959+274947  & 10 39 59  &  +27 49 47 &  4        & 0.0989$\pm$0.0001 &  A          &  -         &                                      \\     
PCG104338+281711  & 10 43 38  &  +28 17 11 &  2+2      &   -       &  C          &  -         &                                      \\
PCG104538+175826  & 10 45 38  &  +17 58 26 &  2        &   -       &  C          &  -         &                                      \\
PCG110941+203320  & 11 09 41  &  +20 33 20 &  3        & 0.1389$\pm$0.0002 &  B          & J1108+2019 & DPOSS II cluster, z=0.139            \\
                  &           &            &           &           &             &            & group at the outskirt of the cluster \\
PCG114233+140738  & 11 42 33  &  +14 07 38 &  3        & 0.1251$\pm$0.0002 &  B          & J1143+1358 & DPOSS II cluster, z=0.120            \\
                  &           &            &           &           &             &            & group at the outskirt of the cluster \\
PCG114333+215356  & 11 43 33  &  +21 53 56 &  4        & 0.1319$\pm$0.0002 &  A          &  -         & SHK 371                              \\ 
PCG121157+134421  & 12 11 57  &  +13 44 21 &  0        &   -       &  C          &  -         &                                      \\
PCG121252+223519  & 12 12 52  &  +22 35 19 &  3        & 0.0850$\pm$0.0002 &  A          &  -         &                                      \\ 
PCG121516+153357  & 12 15 16  &  +15 33 57 &  2        &   -       &  C          &  -         &                                      \\
PCG130157+191511  & 13 01 57  &  +19 15 11 &  3        & 0.0794$\pm$0.0001 &  A          &  -         &                                      \\
PCG130926+155358  & 13 09 26  &  +15 53 58 &  4(3)     & 0.1486$\pm$0.0001 &  A          &  -         &                                      \\
PCG145239+275905  & 14 52 39  &  +27 59 05 &  4(3)     & 0.1255$\pm$0.0001 &  B          & Abell 1984 & group at the outskirt of the cluster \\
                  &           &            &           &           &             &            & z$_{cluster}$=0.124                  \\
                  &           &            &           &           &             &            & SHK 219                              \\ 
PCG154930+275637  & 15 49 30  &  +27 56 37 &  2        &   -       &  C          &  -         &                                      \\
PCG161754+275834  & 16 17 54  &  +27 58 34 &  3        & 0.1259$\pm$0.0001 &  A          &  -         &                                      \\
PCG170458+281834  & 17 04 58  &  +28 18 34 &  2        &   -       &  C          &  -         &                                      \\ \hline
\end{tabular}
\end{table*}

\subsection{{\it Properties of the group galaxies}}

We then analyzed the morphological type of the galaxies in our
candidate compact groups. Lacking good quality images, we adopted the
same criterium as in  Ribeiro et al. (1998), i.e.  we assumed that a
galaxy has a late morphological type if EW(H$\alpha$) $>$
6.0$ \AA$\footnote{We assume here that an emission line has positive
EW}. The EW(H$\alpha$) has been estimated on the wavelength
calibrated spectra using the IRAF task {\it splot}; to separate the
H$\alpha$ line from the two nearby [NII] lines, we used the
{\it deblend} task and the results are shown in Table 4 only for groups of
class A and B.\newline
We assumed that a group can be considered of late morphological type if
at least 50$\%$ of its member galaxies have an EW(H$\alpha$) $>$
6$\AA$, which gives us 5 late type groups from the 11 confirmed
targets, see Table 4. 
Of the 37 galaxies belonging to the confirmed groups, 13 show 
emission in H$\alpha$, for a total of 35$\%$ emitters. No galaxy
shows a velocity dispersion characteristic of a Sy2, but 1 of them is
a starburst galaxy, (EW(H$\alpha$) $\ge$ 50$\AA$, Kennicutt \& Kent,
1983), PCG130157+191151C, and another is an HII galaxy, PCG130157+191151A (see below).
This percentage of emission line objects is equal to what has been
found for Southern Compact Groups and for Hickson groups, for which
the fraction of emission line galaxies is $\sim$ 35$\%$ of the total
(Coziol et al. 2000, 2004).

\begin{table}
\caption{\label{Morphs} Morphological classification of the member galaxies for groups of class A and B.
T$_{spectrum}$ is the morphogical type of the group as defined by the percentage of galaxies
with EW(H$\alpha$)$>$ 6$\AA$}
\begin{tabular}{lllll} \hline 
Group name                & members       & n\_em$^a$ &  T$_{spectrum}$ &  EW(H$\alpha$)      \\ \hline
PCG100355+190454         &  4             & 0         &  early          &        -            \\
PCG101345+194541         &  4             & 0         &  early          &        -            \\
PCG103959+274947         &  4             & 1         &  early          & A: 6.05$\pm$0.01   \\
PCG110941+203320         &  3             & 1         &  early          & B: 12.98$\pm$0.02 \\
PCG114233+140738         &  3             & 1         &  early          & B: 7.60$\pm$0.02 \\
PCG114333+215356         &  4             & 3         &  late           & A: 22.06$\pm$0.04 \\
                         &                &           &                 & B: 6.89$\pm$0.02 \\ 
                         &                &           &                 & C: 8.77$\pm$0.03 \\
PCG121252+223519         &  3             & 2         &  late           & A: 24.45$\pm$0.03 \\ 
                         &                &           &                 & B: 21.64$\pm$0.06 \\
PCG130157+191511         &  3             & 3         &  late           & A: 36.6$\pm$0.2 \\ 
                         &                &           &                 & C: 202$\pm$5 \\ 
                         &                &           &                 & D: 47$\pm$2 \\
PCG130926+155358         &  3             & 1         &  early          & C: 16.60$\pm$0.05 \\
PCG145239+275905         &  3             & 1         &  early          & B: 7.30$\pm$0.08 \\
PCG161754+275834         &  3             & 0         &  early          &        -             \\ \hline
\multicolumn{5}{l}{$^a$ \small number of emitters}
\end{tabular}
\end{table} 
 
\subsection{{\it Individual targets}}

Here we give a brief description of the individual compact groups
selected for this study. The morphological characterisation of the
galaxies is based on their measured EW(H$\alpha$), as stated in the
previous section.

\begin{itemize}
\item{}{\bf PCG100355+190454}: this group is composed of 4 early type
galaxies in a low background density region. The main galaxy of the
group falls very close to a bright star, which could not be separated
from the galaxy on the DPOSS II plate. It is the classical example of
a group failing the selection criteria and contaminating the
sample. If the deblending algorithm had succeeded in separating the
galaxy from the nearby bright star, the group would not have been
included in the final sample, as none of its member galaxies are
brighter than 17 mag in r.  Even if classified as a Class A group, it
should be excluded from the final sample.

\item{}{\bf PCG101345+194541}: this group deserves special mention
as five additional galaxies were observed at the same time as the candidate
group members, for a total of nine observed galaxies, all of them
are within 1000 \kms of the mean group velocity. Checking
the DSS image, we found that the group actually coincides with a
cluster, Abell 0952, whose redshift was measured by Miller et
al. (2002), and we can now confirm its location at z = 0.1121$\pm$0.0002. If we
re-measure the velocity dispersion, the crossing time, and the mass and
mass-to-light ratio taking all nine galaxies into account, we find
substantially different values from those measured using only four
galaxies (see next sections): log($\sigma_{r}$) = 2.76;
log($H_\mathrm{o}t_\mathrm{c}$) = -2.03, M = 4.11 x 10$^{13}$
$M_\mathrm{\sun}$ , L = 3.23 x 10$^{11}$ $L_\mathrm{\sun}$, M/L =
127. These numbers are indeed more characteristic of a cluster than a
compact group; however, they are slightly different from those found by
Miller et al. with 10 galaxies. He quotes a velocity dispersion of 512
\kms against 582 \kms measured by us. This might be explained by the
fact that our observed galaxies form an elongated structure on the
sky, while the object observed by the other group are distributed more
homogeneously around the centre of the cluster. In fact repeating once
more the calculation taking all the observed 19 galaxies
(ours+Miller) into account,  we find a value of 535 \kms, in good agreement, within
the errors, with Miller's estimate. The target has been classified as
class B {\it group}. \newline
It should be
noted that this group is not in the final list of DPOSS
compact groups candidates in Carvalho et al. (2005). It is located on a
plate whose quality was not good enough to ensure reliable star
galaxy separation (explaining a posteriori why such object entered our
sample). 

\item{}{\bf PCG103349+225324}: two pairs of galaxies at discordant
redshifts.  The brightest galaxy of the nearest pair at z = 0.063$\pm$0.0002,
which is also the main galaxy of the group, shows a disturbed
morphology, with asymmetric arms and it is likely to be interacting
with galaxy D. The more distant pair, at z = 0.1106$\pm$0.0001, doesn't show any
emission features, but disks are clearly visible in both galaxies.
The group has been classified as class C.

\item{}{\bf PCG103959+274947}: this group is composed of 4 concordant
galaxies, of which only one, the brightest of the group, shows clear
spiral arms superimposed on a disk and H$\alpha$+[NII] emission
lines. It looks like the classical example of a compact group and
has been classified as A.

\item{}{\bf PCG104338+281711}: two pairs of galaxies at discordant
redshifts.  Both galaxies of the closest pair at z = 0.0810$\pm$0.0001, show a
disk and H$\alpha$ emission, while the more distant ones at z = 0.2450$\pm$0.0003
shows no emission lines. The group has been classified as class C.

\item{}{\bf PCG104538+175826}: A disk is clearly visible in galaxies A
and B, but A has a different redshift from the other galaxies in the group. 
The disk in B looks asymmetric, and the galaxy is at the same
redshift, z = 0.1060$\pm$0.0001, as C with which it might be interacting. 
The group has been classified as class C.

\item{}{\bf PCG110941+203320}: a group composed of 1 late type galaxies
and two early types; the emission line galaxy, B, has an
H$\alpha$ emission with lower flux than the [NII], probably due to underlying
absorption from old stellar population. It is close to the DPOSS
cluster J1108+2019 and classified as type B.

\item{}{\bf PCG114233+140738}: Galaxy A is the only one with a clearly
visible disk in the acquisition image, but it is also the discordant
member of the group at z = 0.0990$\pm$0.0001. Galaxy B shows some emission
lines and is classified as late type galaxy, but the other two
galaxies are early type galaxies. Since galaxy B has a magnitude of
16.7, the group doesn't fail the selection criterion that requires 
the brightest galaxy to have a magnitude between 16 $\le$ r $\le$ 17. Close to the
DPOSS II cluster J1143+1358 this nice triplet of early type galaxies
has been classified as B.

\item{}{\bf PCG114333+215356}: a group dominated by emission line
galaxies.  Galaxy B seems to be interacting with another object, but
the resolution of the acquisition image doesn't allow better
investigation. Its H$\alpha$ profile shows a narrow peak superimposed
on a larger line, probably coming from the other galaxy, which is
almost superimposed on galaxy B. A fifth galaxy, E, is also associated
with this group, but its faintness did not allow us to obtain a good
spectrum. This group was already known in the literature as SHK 371
(Stoll et al., 1997), but its redshift was measured here for the
first time. The group belongs to Class A.

\item{}{\bf PCG121157+134421}: this group is the only one of our
sample with 4 discordant galaxies. Galaxy A, the brightest one, has a
clear disk-like morphology similar to galaxy D, which is, however, at
a different redshift. The group has been classified as class C.

\item{}{\bf PCG121252+223519}: this group turned out to be a triplet,
with galaxy D a star. The triplet is composed of two late type
spiral galaxies, and an early type galaxy, C.
Spiral arms are clearly visible in galaxy A, which shows also significant emission 
in H$\alpha$ (EW = 24.45$\pm$ 0.03 $\AA$), while a disk is evident
in galaxy B. The group has been classified as class A.

\item{}{\bf PCG121516+153357}: nice candidate group of four galaxies,
two of which show emission lines. Unfortunately it is composed of
one pair (A, D) at z = 0.0980$\pm$0.0001 and two discordant
galaxies. Interestingly enough, galaxy B has a velocity difference of
$\sim$ 1970 \kms from the AD pair. This is too much for our
selection criteria, so the group was not confirmed.  However, a DPOSS 
cluster, J121528+153533 (Gal et al.,
2003), with a quoted photometric redshift of 0.147, is found within
the 15$\arcmin$ radius used for our proximity search. The group has been classified as class C.

\item{}{\bf PCG130157+191511}: an emission line triplet composed of late
type spirals and galaxy B, which is in fact a star. Galaxies A and C
show disturbed morphology with asymmetric arms and a hint of a tidal
tail.  Galaxy C shows strong emission lines characteristics of a
starburst galaxy with an EW(H$\alpha$)= 202$\pm$5$\AA$, while
galaxies A and D are also strong emitters. This lead us to try a
comparison with the most active group in the Hickson sample, HCG 16
(Ribeiro et al., 1996); indeed, the line ratios ([OIII]5007/H$\beta$)
and ([NII]6584/H$\alpha$) of the C and D galaxies are characteristics
of starburst nuclear galaxies, while A is located between HII galaxies
and SBNG. If we extract the spectra using
an aperture of 1$\arcsec$, i.e. 1.58 kpc, all the three galaxies show
line ratios characteristics of starburst nuclear galaxies (SBNGs). 
The group has been classified as class A.

\item{}{\bf PCG130926+155358}: again, galaxy B turned out to be a
star; one more galaxy which has been included in the slit happened to be at the same
redshift of the candidate member galaxies, but it was not taken into
account in the calculation of the group parameters. This group is
dominated by early type galaxies, since only galaxy C shows emission
in H$\alpha$ and [NII]. The group has been classified as class A.

\item{}{\bf PCG145239+275905}: a group composed of 3 galaxies, one of
them showing emission lines. It was already known in the literature as
SHK 219 and is at the edge of the Abell cluster 1984. As for SHK
371, the redshift of this group was measured here for the first
time. This is a class B group, but it should be noted that this group fails the 
selection criterion of having the r magnitude of the brightest galaxy
between 16 and 17, so it should be excluded from the final sample.

\item{}{\bf PCG154930+275637}: Galaxies A, C, and D have a diffuse
appearance, and C and D seem to be interacting in the acquisition
image. Unfortunately only galaxies A and D are concordant at z =
0.0760$\pm$0.0002. The group has been classified as class C.

\item{}{\bf PCG161754+275834}: a group dominated by early type galaxies
with no emission line detected. The spectrum of galaxy D is
contaminated by strong moon reflection and could not be used. 
The group has been classified as class A.

\item{}{\bf PCG170458+281834}: This group is composed of an
interacting pair (A,C) at z = 0.1640$\pm$0.0002, while galaxy B is a star and
galaxy C is a foreground galaxy.  All the candidates look like early
type galaxies (E/S0) in the acquisition image.   The group has been classified as class C.
\end{itemize}
\vspace{1.0cm}

In Fig. 2, which is available in the electronic version of the paper, 
we show the acquisition images of the concordant groups
(class A and B), while in Fig. 3, also available in the electronic
version only,  the spectra of their member
galaxies are shown.
At this point one might wonder about residual contamination of the
sample by mis-classified galaxies, i.e. stars. If we assume that our
sample is representative of the full compact group DPOSS II survey, we
can say that, out of 72 galaxies, we find that 4 have been
misidentified and are stars, which means that the residual
contamination from incorrect star/galaxy separation is 5.5$\%$.
This rate is slightly lower than the quoted error of 9$\%$
in Odewahn et al. (2004); however the authors
admit that the 9$\%$ is most likely an overestimation of the real error.
Indeed, when looking at their Table 2, it appears that our percentage
of wrong star-galaxy separation is comparable to that
found by Odewahn et al. for a similar magnitude range.\newline

In the following sections we discuss only the properties of the
groups classified as A and B.

\subsection{{\it Internal dynamics and mass estimates}}

In Fig. 4 we show the distribution of the velocity dispersion for
the observed groups, which is perfectly consistent with that of
HCGs and Southern Compact Groups (hereafter SCGs, Iovino,
2000), while it is smaller than the average value found for loose
groups of galaxies (see for example Eke et al., 2004).
The base 10 logarithm of the values
measured for the confirmed groups are shown in column 3 of Table
5.\newline

\setcounter{figure}{3}
\begin{figure}[h]
\resizebox{\hsize}{!}{\includegraphics{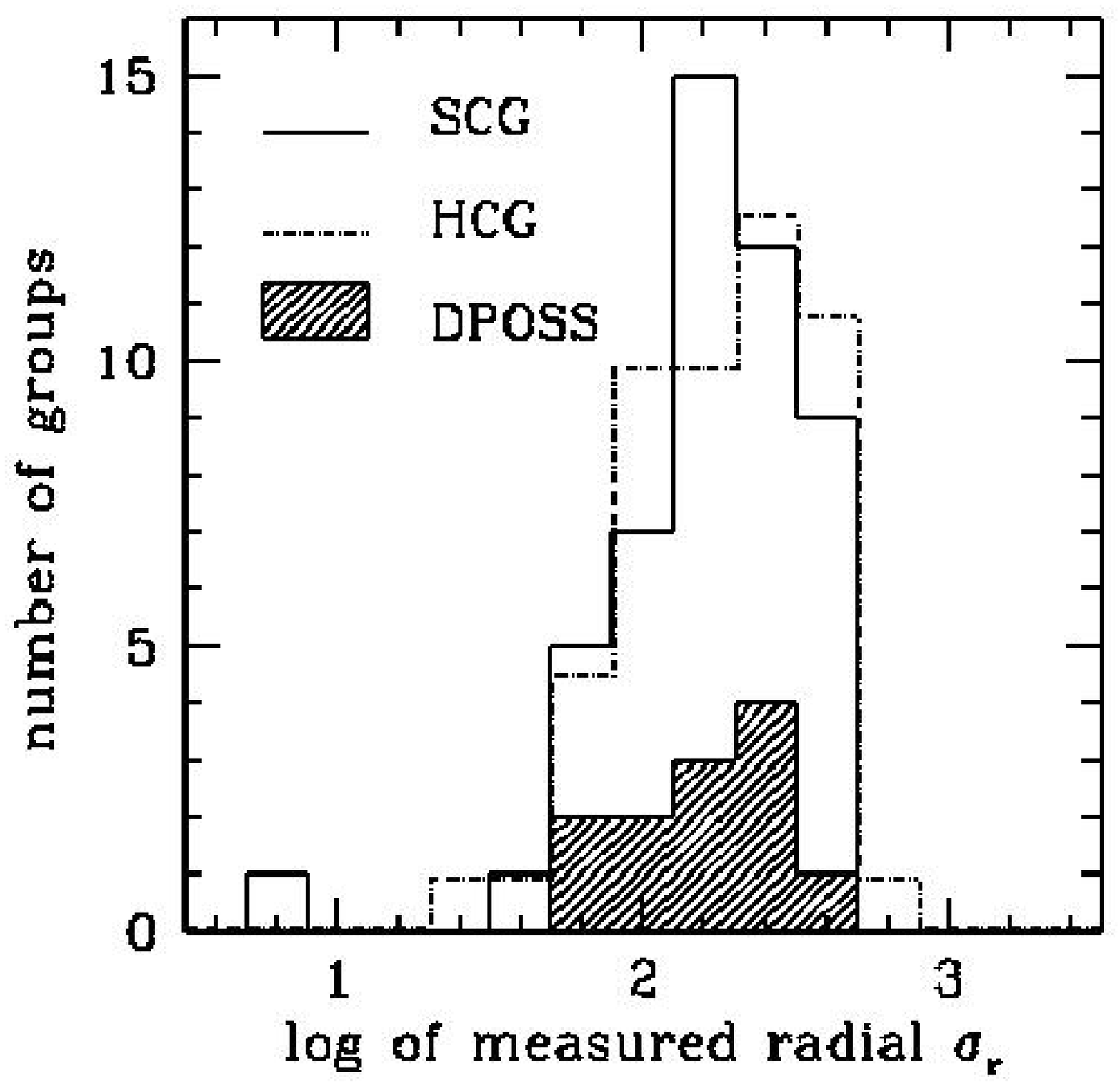}}
\caption{Distribution of the observed velocity
dispersion for the DPOSS groups (hatched area), 
Southern Compact Groups survey (SCGs, strong line) 
and Hickson Compact Groups (HCGs, light line)}
\label{Fig. 4}
\end{figure}

To have an idea of how our groups compare to HCGs and
to SCGs, we netx estimated the crossing time, defined as (Hickson et al., 1992):

\begin{equation}
t_\mathrm{c} =  \frac{4}{\pi} 
                \frac{R}{\sigma_{3D}}
\end{equation}

where R is median of galaxy-galaxy separation and $\sigma_{3D}$ is 
the three dimensional velocity dispersion, defined as in Hickson et al.(1992). The
logarithm of the crossing time measured for our groups are
listed in column 5 of Table 5, while the distribution of crossing times
is shown in Fig. 5.  The median value of $\mathrm{t_{c}}$
is 0.018 $H_\mathrm{o}^{-1}$, in good agreement with what was measured for HCGs,
0.016 $H_\mathrm{o}^{-1}$.\newline

\setcounter{figure}{4}
\begin{figure}[h]
\resizebox{\hsize}{!}{\includegraphics{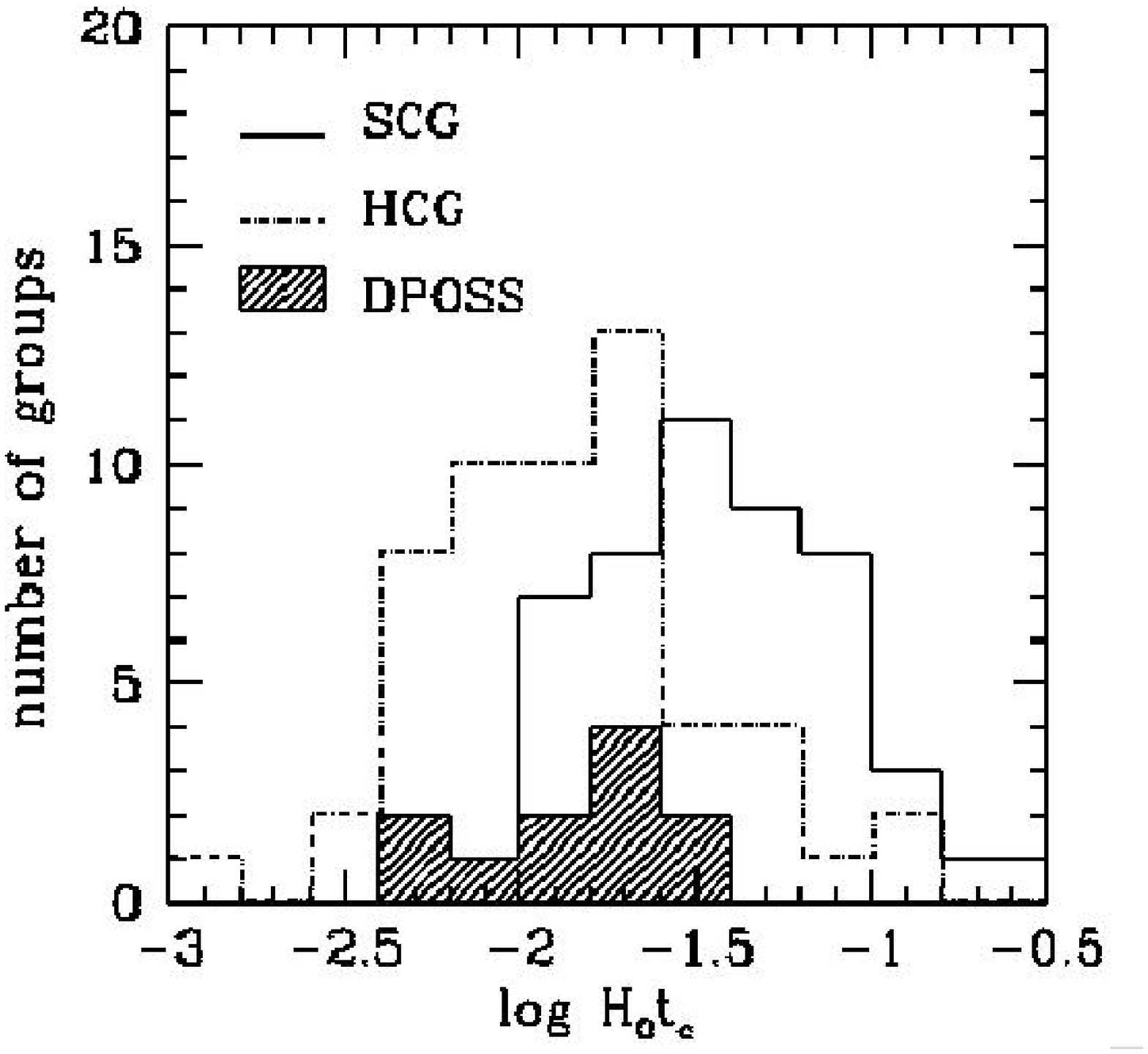}}
\caption{Distribution of dimensionless crossing time for
the confirmed DPOSS groups (hatched area), compared to that
obtained for HCGs (light line) and for
SCGs (heavy line).}
\end{figure}

Finally, we provide an estimate of the group mass and the corresponding
mass-to-light ratio in the Gunn {\it r} filter.
To estimate the mass of the groups, we assume that we can use the
virial theorem, so that the expression for the mass is:

\begin{equation}
\mathrm{M_{V}}  = \frac{3\pi\mathrm{N}}{2\mathrm{G}}
                  \frac{\mathrm{\Sigma_{i}\sigma_{ri}^2}}{\mathrm{\Sigma_{i<j}1/R_{ij}}}
\end{equation}

where $R_{ij}$ is the projected separation between galaxies i and j,
here assumed to be the median length of the two-dimensional
galaxy-galaxy separation vector, corrected for cosmological effects. N is
the number of concordant galaxies in the system, and $\sigma_{ri}^2$ the velocity component
along the line of sight of the galaxy {\it i} with respect to the centre
of mass of the group.
As observed by Heinsler et al. (1985) and by Perea et al. (1990), the
use of the virial theorem produces the best mass estimates, provided
that there are no interlopers or projection effects.  In case one of
these two effects is present,  the current values can be considered an
upper limit to the real mass.
Equation 2 is valid only under the
assumption of spherical symmetry and isotropy.  Another mass estimate
is given by the projected mass estimator, defined as:

\begin{equation}
\mathrm{M_{P}} = \frac{f_{P}}{GN} 
                 {\mathrm{\Sigma_{i}\sigma_{ri}^2}\mathrm{R_{i}}}
\end{equation}

where $R_{i}$ is the projected separation from the centroid of the system,
and f$_{P}$ is a numerical factor depending on the distribution of the orbits
around the centre of mass of the system.
Assuming a spherically symmetric system for which the Jean's hydrostatic equilibrium
applies, we can express f$_{P}$ in  an explicit form (Perea et al., 1990). 
Since we lacked information
about the orbit eccentricities, we assumed isotropic orbits and the expression for
M$_{P}$ became:

\begin{equation}
\mathrm{M_{P}} = \frac{64}{2\pi G}
                 {<\sigma_{r}^2 R>}
\end{equation}
 
where again R is the median length of the two-dimensional
galaxy-galaxy separation vector.\newline
We calculated both quantities, M$_{V}$ and
M$_{P}$, and found that M$_{P}$ is a few percent smaller than M$_{V}$, but
of a comparable order of magnitude. The mass estimate we report in column 6 of
Table 5 is the average of the two estimators. Errors on the estimate of the mass
are largely dominated by the difference between the two estimators, which has been
assumed as total error.\newline

To estimate the luminosities, we use the {\it r} band magnitudes of the groups 
obtained by summing up all the flux of the member galaxies as measured on the
calibrated DPOSS II plates and also published in Iovino et al., 2003: these are corrected for galactic
extinction, and for k-correction, and re-scaled to absolute
value.
Two different k-corrections have been defined: one for early type 
galaxies (E-Sa) and another for spirals (Sbc): the EW(H$\alpha$) has been used to discriminate
in this sense.\newline

As a reference for the solar magnitude, we used the paper by Jorgensen
(1994). From stars with a $(B-V) \sim 0.65$ (i.e. the same colour of
the sun) it is possible to estimate a colour index $(r-R) \sim 0.354$. 
Taking $M_\mathrm{R,\sun}$ = 4.42
(Binney \& Merrifield, 2001), we have $M_\mathrm{r,\sun}$ = 4.77.
The values for the light and the mass-to-light 
ratio are shown in columns 7 and 8 of Table 5 respectively.
Errors on the luminosity were estimated by assuming the maximum error
on the photometric calibration of DPOSS plates, i.e. an error of 0.19
magnitudes for an {\it r} magnitude of 19 (Gal et al. 2004).\newline

The median value of the $M/L_{r}$ in the sample is 47.
It should be noted, however, that Hickson used B-band luminosity, not r. If 
we use the B band luminosity,
assuming the transformation (Windhorst et al., 1991),

\begin{equation}
B  =  g+0.51+0.60\times(g-r)
\end{equation}

and if we take $M_\mathrm{B,\sun}$ = 5.48, we found that the median $M/L_{B}$ is 
92 {\it h}, about 50$\%$ bigger than what has been estimated for HCGs. We re-scaled our values to
the $H_{0}$ used by Hickson, 100 \kms, with a value of {\it h}=0.67, to allow easier
comparison. The {\it (g-r)} colours of the galaxies come from Iovino et al. (2003).
Both values are
lower than those measured for loose groups, between 200 and
400 {\it h}, but still higher than the value measured for single
galaxies in HCGs, 7 {\it h}, see Rubin et al. (1991).

\begin{table*}
\caption{\label{group_prop} Group dynamical properties. $\sigma_{r}$, R, and $H_\mathrm{o}t_\mathrm{c}$ are
expressed in logarithmic values; mass,
luminosity,and mass-to-light ratio are given in solar units. The symbols follow Hickson et al. (1992)}
\begin{tabular}{c c c c c c c c} \hline
Group name   &   Scale     & $\sigma_{r}$ & log(R) & $H_\mathrm{o}t_\mathrm{c}$ & M                    & L                          & M/L   \\ 
             &(kpc/$\arcmin$) & (\kms)   & (kpc)  &                             & $M_\mathrm{\sun}$    & $L_\mathrm{\sun}$           &       \\ \hline
PCG100355+190454 &  123.44     & 1.89          & 1.71   & -1.478                & $(1.4\pm0.3)\times 10^{12}$ & $(1.4\pm0.1)\times 10^{11}$   & 9  \\
PCG101345+194541 &  136.57     & 2.49          & 1.67   & -2.133                & $(1.3\pm0.2)\times 10^{13}$ & $(1.13\pm0.09)\times 10^{11}$ & 108   \\
PCG103959+274947 &  114.70     & 2.36          & 1.46   & -2.203                & $(4.5\pm0.8)\times 10^{12}$ & $(7.5\pm0.6)\times 10^{10}$   & 55 \\
PCG110941+203320 &  155.3      & 2.22          & 1.79   & -1.733                & $(6\pm2)\times 10^{12}$     & $(1.11\pm0.08)\times 10^{11}$ & 48 \\
PCG114233+140738 &  142.06     & 2.49          & 1.48   & -2.315                & $(1.0\pm0.3)\times 10^{13}$ & $(1.4\pm0.1)\times 10^{11}$   & 63 \\
PCG114333+215356 &  147.31     & 1.91          & 1.61   & -1.606                & $(1.3\pm0.2)\times 10^{12}$ & $(1.8\pm0.1)\times 10^{11}$   & 6 \\
PCG121252+223519 &  101.06     & 2.01          & 1.47   & -1.845                & $(1.6\pm0.4)\times 10^{12}$ & $(1.05\pm0.08)\times 10^{11}$ & 13 \\
PCG130157+191511 &   95.02     & 1.91          & 1.50   & -1.715                & $(7\pm2)\times 10^{11}$     & $(6.3\pm0.4)\times 10^{10}$   & 10 \\
PCG130926+155358 &  162.78     & 2.21          & 1.97   & -1.552                & $(8\pm2)\times 10^{12}$     & $(1.5\pm0.1)\times 10^{11}$   & 47 \\
PCG145239+275905 &  142.45     & 2.39          & 1.77   & -1.920                & $(1.2\pm0.3)\times 10^{13}$ & $(6.9\pm0.5)\times 10^{10}$   & 149 \\
PCG161754+275834 &  141.46     & 2.13          & 1.73   & -1.703                & $(3.4\pm0.9)\times 10^{12}$ & $(1.24\pm0.09)\times 10^{11}$  & 24 \\ \hline
\end{tabular}
\end{table*} 

\subsection{{\it Radius distribution}}
We can now wonder whether we are indeed looking at compact groups
or looser structures. In the first case, we would expect a peak
in the distribution around $\sim$ 50 kpc, while in the second
a wider distribution is expected.\newline

In Fig. 6, we show the distribution of the physical radius for our
confirmed compact groups (classes A and B).  We observe a narrow
distribution of physical sizes around $\sim$ 50 kpc, with the values
greater than 100 kpc given by the very elongated group PCG
130926+155358.
This very well agrees with the characteristics physical scale
for compact groups, 50 Kpc, hence we can be confident within our
small number statistics that we are indeed observing compact groups.

\setcounter{figure}{5}
\begin{figure}
\centering
\includegraphics[angle=-90,width=9cm]{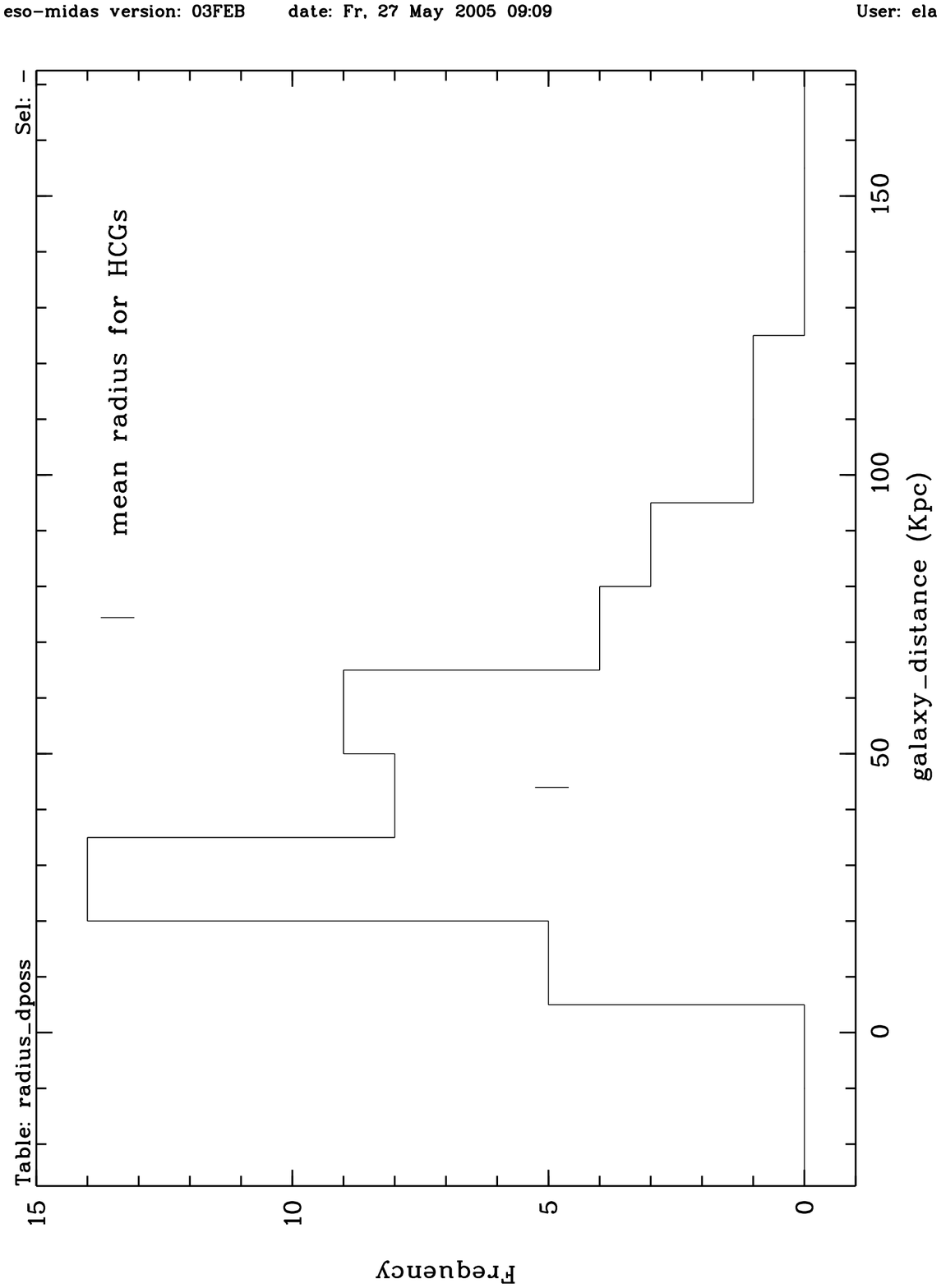}
\caption{Distribution of the galaxy-galaxy distance for class A and B
groups. The peak of the distribution is consistent with the
characteristic physical scale of compact groups, $\sim$ 50 kpc}
\label{Figure 6}
\end{figure}

\section{Discussion}

Observation and analysis of this
small sample of 18 candidate groups have shown
several important results of the DPOSS survey:
\begin{itemize}
\item{} Despite the (extremely) low number statistics, 
the algorithm that identifies compact 
structures on the DPOSS survey has a success rate
of 60$\%$ in identifying dense bound structures on the sky even at
an average redshift of 0.2; if we consider only isolated compact
groups, those objects with 4 spectroscopically concordant galaxies and no neighbour, this rate
drops to 22$\%$.
\item{} There is only one candidate group with 4 discordant galaxies:
the worst cases, class C groups, were otherwise always composed of
at least one pair of galaxies at a similar redshift.
\item{} The masses of the groups are higher than
those found for HCGs, but the median value of 4.5 x 10$^{12}$ $M_\mathrm{\sun}$
for our sample and 1.5 x 10$^{12}$ $M_\mathrm{\sun}$ for the HCGs are comparable.
\item{} We are able to identify really compact configurations on the
sky, with an average radius of $\sim$ 50 kpc, which is a scale that is usually
missed in the current redshift surveys, mainly due to the
fiber separation limit.
\item{} About 27$\%$ of the confirmed groups are late
type systems.
\item{} Interestingly enough, the most active group of our sample,
PCG 130157+191511, contains three star-forming galaxies in it; the only group 
showing a similar
activity in the Hickson's catalog is HCG 16.
\end{itemize}

As for the first two points, we can ask how this survey work compared 
with other higher redshift surveys,
namely the Las Campanas redshift survey (Allam \& Tucker, 1999, 2000;
Tucker et al., 2000, hereafter LCRS), the Sloan (Lee et al., 2004), and
the 2dF galaxy redshift survey.
In the LCRS catalogue it is possible to find two group catalogs: one
of compact groups (Allam \& Tucker) and another of loose groups
(Tucker et al.). Let us then focus for the moment only on the compact
groups catalogue. This catalogue is seriously affected by
the fact that the fibers used in the survey have a minimum separation
on the sky of 55$\arcsec$, so that it is not possible to measure the
individual redshifts of candidate group members whose separation
on the sky is less than 55$\arcsec$. As a consequence,
of the 76 candidate groups of the catalog, only one has measured
redshift for all its members, 23 groups have measured redshifts
for two member galaxies, while the rest have just one.\newline

The Sloan survey data do not allow redshift determination for
individual group members, again due to the large separation on the
sky of the fibers used to carry on the survey: 60$\arcsec$, i.e.
80h$^{-1}$ kpc at z $\sim$ 0.1.  Basically, these
data are affected by the same problem as LCRS. The identification
of a group is made by assigning the same redshift of the brightest
galaxy to all galaxies within
the fiber limitation circle, or, in the best cases, a group is identified by
measuring only two redshifts.\newline 

Since we have shown here that only one candidate group out of 18
has all its members with discordant redshifts, the
technique used by the Sloan cannot uniquely identify a compact group.
In a similar
way, the 2dF survey is affected by limitation due to the fiber
diameter; however, many fields have been observed more than once
with the fibers in different positions, thus reducing the problem.
However, the catalogue created by Merchan
\& Zandivarez (2002) covers a wide range in redshift (0.003 to 0.25),
and the mass
range and crossing times also include objects that are much more like
loose groups than to compact groups.\newline

Finally, one might object that an average redshift of 0.2 is not
so very high in an epoch where discovery of z = 3 targets is common.
While this is true, it is important to keep in mind that the projected size
of a typical compact group at z $\sim$ 0.01, 1$\arcmin$
corresponds to 13 kpc on the sky, i.e. to a fourth of the typical radius
of a compact group (50 kpc). At z $\sim$ 0.1, 1$\arcmin$ is already twice
the size of a compact group, and at z $\sim$ 1 a compact group is spread
over an area of $<$ 20$\arcsec$. To this one must add the increasing fore/background
contamination of other objects, making the task of identifying and
studying compact groups
at higher redshift quite a challenging one.\newline

Summing up, we can reasonably conclude that, of the existing
surveys of compact groups at intermediate $z$, the DPOSS II survey and
its spectroscopic follow-up will indeed provide, once completed, an unprecedented
database for the study of compact groups outside the local universe.
The other important question is what the properties of confirmed
DPOSS groups tell us. Having only 11 targets makes it difficult to draw firm conclusions.
However, we can point out what we would expect
if the majority of the DPOSS groups were to show the same characteristics.
The fraction of late type galaxies (f$_{s}$ = 0.35) and the crossing times
($H_\mathrm{o}t_\mathrm{c}$ = 0.018) are similar to those measured for Hickson Compact Groups
(f$_{s}$ = 0.49; $H_\mathrm{o}t_\mathrm{c}$ = 0.016), while other group characteristics, like mass,
and velocity dispersion are also very similar. \newline
With these results in hand, we might wonder whether we
have find an answer to our three questions: with increasing
redshift, do we find:
\begin{itemize}
\item{}a high number of interacting groups?
\item{}an increase in the number of active groups?
\item{}a change in the average group's physical parameters?
\end{itemize}
The first two questions are still open because, while we do
find a strong star-forming group among 11 targets, we feel that the
statistics are still too poor to extrapolate
this finding to the full sample. Ongoing observations at La Silla
on the southern part of the sample will help to
improve the statistics.
For the last question, the answer
seems to be no, because our results point toward no
evolution for compact groups up to z = 0.12, which is the median redshift of
our observed sample.\newline

This indeed looks puzzling, since a redshift of z = 0.12 corresponds to a look-back
time of 1.56 Gyr, while our compact groups should have dissolved in 0.21
Gyrs, given their median crossing time of $t_\mathrm{c}$ = 0.018$H_\mathrm{o}^{-1}$.
This could mean that either we are looking at a different set of groups
from the one we observe in the nearby Universe, e.g. groups that have already merged by
the present time, leaving a field elliptical or a substructure within a cluster,
or there is something that stabilizes the groups extending their lifetime.\newline

This seems to agree well with
numerical models that favour an early formation of a common, massive
halo within which individual galaxies form (Gomez-Flechoso \& Dominguez-Tenreiro, 2001).
This model, unlike the one proposed by Atahnassoula \& Makino (1997),
predicts a central concentration for the common halo, in agreement
with the observations. Moreover, according to this model, the galaxy interactions
perturb the global halo potential and become significant in
changing the group only if they are comparable to the global
field of force of the halo. This scenario seems consistent with
previous findings that individual galaxy properties within the
group do not correlate with the group global properties.
It must be kept in mind, however, that active groups with a low velocity dispersion, 
like those present in the Southern Compact Group survey,
are still not explained well by this scenario, so that a more detailed
investigation is needed to understand the nature of
compact groups completely.

\section{Conclusions}
Our pilot study of compact groups at intermediate redshift has
shown that the confirmed candidates have very similar properties
to those observed for Hickson Compact Groups and that
no significant evolution can be detected up to z $\sim$ 0.12.
This finding agrees with models predicting an early formation
of a massive, common halo, within which the individual galaxies
form and evolve. Such models, however, are still unable to explain
the low velocity dispersion, high activity level groups found
in the nearby universe.\newline
Our results suggest that
the DPOSS sample, once full redshift information for its members
becomes available, will provide a reference sample for studying the
properties of compact groups beyond the local universe.

\begin{acknowledgements}

It is a pleasure to thank the La Silla
Science Operation team for their help during the 
observations and daytime support.
RRdC would like to thank Roy Gal
for several insightful discussions on the subject.\newline
This research made use of the NASA/IPAC Extragalactic
Database (NED), which is operated by the Jet Propulsion
Laboratory, California Institute of Technology, under
contract with the National Aeronautics and Space Administration.
\end{acknowledgements}

\Online
\setcounter{table}{1}
\begin{longtable}{c c c c c c}
\caption{\label{dposs_gals} Observed compact groups candidates from the DPOSS survey. Column 5 lists
the kind of spectrum for each galaxy: {\it abs} means a spectrum which is dominated by absorption lines; 
{\it em} means a spectrum dominated by emission lines and {\it mix} a spectrum
where both emission and absorption lines are present. The most significant emission lines
have been listed for the confirmed member galaxies.} \\
\hline\hline
Galaxy name          &  $\alpha$   & $\delta$      &   cz          & line & emission lines                                         \\ 
                     & (J2000.0)   & (J2000.0)     & (\kms) &      &                                                               \\ \hline
PCG100355+190454A & 10 03 54.22 & +19 05  2.11 & 32234$\pm$ 75 & abs  &  -                                                         \\ 
PCG100355+190454B & 10 03 56.92 & +19 05  2.11 & 32494$\pm$ 56 & abs  &  -                                                         \\
PCG100355+190454C & 10 03 55.70 & +19 04 55.09 & 32294$\pm$ 42 & abs  &  -                                                         \\
PCG100355+190454D & 10 03 53.61 & +19 04 47.21 & 32355$\pm$ 42 & abs  &  -                                                         \\ \hline
PCG101345+194541A & 10 13 45.16 & +19 45 44.28 & 33923$\pm$ 40 & abs  &  -                                                         \\
PCG101345+194541B & 10 13 46.87 & +19 45 30.71 & 33254$\pm$ 41 & abs  &  -                                                         \\
PCG101345+194541C & 10 13 44.87 & +19 45 51.52 & 33378$\pm$ 32 & abs  &  -                                                         \\
PCG101345+194541D & 10 13 45.84 & +19 45 37.61 & 34033$\pm$ 82 & abs  &  -                                                         \\ \hline
PCG103349+225324A & 10 33 48.96 & +22 53 44.60 & 18775$\pm$ 68 & em   & [OII],H$\beta$,[OIII],H$\alpha$,[NII],[SII]                \\
PCG103349+225324B & 10 33 49.50 & +22 53 14.43 & 32949$\pm$ 59 & abs  &  -                                                         \\
PCG103349+225324C & 10 33 50.26 & +22 53 15.68 & 33413$\pm$ 56 & abs  &  -                                                         \\
PCG103349+225324D & 10 33 47.75 & +22 53 34.15 & 18869$\pm$ 70 & em   & H$\alpha$,[NII],[SII]                                      \\ \hline
PCG103959+274947A & 10 39 59.23 & +27 49 35.72 & 29900$\pm$ 20 & em   & H$\alpha$,[NII]                                            \\
PCG103959+274947B & 10 39 58.52 & +27 49 44.98 & 29863$\pm$ 32 & abs  &  -                                                         \\
PCG103959+274947C & 10 39 59.13 & +27 49 59.48 & 30503$\pm$ 28 & abs  &  -                                                         \\  
PCG103959+274947D & 10 39 58.90 & +27 49 41.88 & 29764$\pm$ 52 & abs  &  -                                                         \\ \hline
PCG104338+281711A & 10 43 38.19 & +28 17 21.23 & 23339$\pm$ 29 & em   & H$\alpha$,[NII],[SII]                                      \\
PCG104338+281711B & 10 43 38.97 & +28 17 13.09 & 25318$\pm$ 85 & em   & H$\alpha$,[NII],[SII]                                      \\
PCG104338+281711C & 10 43 39.40 & +28 17 05.49 & 73583$\pm$ 69 & abs  &  -                                                         \\
PCG104338+281711D & 10 43 37.83 & +28 17 06.68 & 73525$\pm$ 77 & abs  &  -                                                         \\ \hline
PCG104538+175826A & 10 45 39.91 & +17 58 31.48 & 21687$\pm$ 36 & abs  &  -                                                         \\
PCG104538+175826B & 10 45 38.98 & +17 58  7.86 & 31751$\pm$ 50 & abs  &  -                                                         \\
PCG104538+175826C & 10 45 37.78 & +17 58 17.55 & 31654$\pm$ 49 & abs  &  -                                                         \\
PCG104538+175826D & 10 45 37.90 & +17 58 45.16 & 39071$\pm$ 62 & abs  &  -                                                         \\ \hline
PCG110941+203320A & 11 09 42.65 & +20 33 29.67 & 34956$\pm$ 60 & em   & H$\alpha$,[NII],[SII]                                      \\ 
PCG110941+203320B & 11 09 39.79 & +20 33 11.20 & 41593$\pm$ 60 & em   & H$\alpha$,[NII]                                            \\
PCG110941+203320C & 11 09 41.06 & +20 33 35.53 & 41511$\pm$ 55 & abs  &  -                                                         \\ 
PCG110941+203320D & 11 09 40.29 & +20 33 13.28 & 42005$\pm$ 60 & abs  &  -                                                         \\ \hline
PCG114233+140738A & 11 42 33.84 & +14 07 51.56 & 29860$\pm$ 37 & abs  &  -                                                         \\
PCG114233+140738B & 11 42 32.39 & +14 07 25.61 & 37285$\pm$ 90 & em   & [OII],H$\alpha$,[NII]                                      \\
PCG114233+140738C & 11 42 33.78 & +14 07 27.09 & 37362$\pm$ 40 & abs  &  -                                                         \\
PCG114233+140738D & 11 42 32.83 & +14 07 28.27 & 38136$\pm$ 45 & abs  &  -                                                         \\  \hline
PCG114333+215356A & 11 43 33.15 & +21 53 50.32 & 39806$\pm$ 20 & em   & [OII],H$\beta$,[OIII],[OI], H$\alpha$,[NII]                \\
PCG114333+215356B & 11 43 32.87 & +21 54 05.44 & 39586$\pm$ 40 & em   & H$\alpha$,[NII]                                            \\ 
PCG114333+215356C & 11 43 32.93 & +21 53 56.37 & 39558$\pm$ 23 & em   & [OII],H$\alpha$,[NII]                                      \\ 
PCG114333+215356D & 11 43 33.53 & +21 54 21.46 & 39531$\pm$ 72 & abs  &  -                                                         \\ \hline
PCG121157+134421A & 12 11 58.54 & +13 44 29.57 & 24217$\pm$ 15 & mix  & H$\alpha$,[NII]                                            \\
PCG121157+134421B & 12 11 57.30 & +13 44 13.24 & 52149$\pm$ 96 & abs  &  -                                                         \\
PCG121157+134421C & 12 11 57.78 & +13 44 16.33 & 20574$\pm$ 98 & abs  &  -                                                         \\
PCG121157+134421D & 12 11 58.69 & +13 44 19.54 & 87724$\pm$ 100 & abs &  -                                                         \\ \hline
PCG121252+223519A & 12 12 53.31 & +22 35 26.55 & 25502$\pm$ 74 & em   & [OIII],[OI], H$\alpha$,[NII]                               \\
PCG121252+223519B & 12 12 51.93 & +22 35 30.08 & 25457$\pm$ 80 & em   & H$\alpha$,[NII]                                            \\ 
PCG121252+223519C & 12 12 52.42 & +22 35 14.28 & 25770$\pm$ 57 & abs  &  -                                                         \\ 
PCG121252+223519D & 12 12 52.55 & +22 35 06.94 & STAR!         &  -   &  -                                                         \\ \hline
PCG121516+153357A & 12 15 16.30 & +15 34 25.29 & 29456$\pm$ 47 & abs  &  -                                                         \\
PCG121516+153357B & 12 15 15.07 & +15 33 50.91 & 31324$\pm$ 50 & abs  &  -                                                         \\
PCG121516+153357C & 12 15 14.44 & +15 33 57.85 & 24477$\pm$ 50 & mix  & H$\alpha$,[NII]                                            \\
PCG121516+153357D & 12 15 15.71 & +15 33 34.89 & 29535$\pm$ 47 & mix  & H$\alpha$,[NII],[SII]                                      \\ \hline
PCG130157+191511A & 13 01 56.80 & +19 14 54.93 & 23750$\pm$ 30 & em   & [OII],H$\beta$,[OIII],[OI],H$\alpha$,[NII],[SII]           \\ 
PCG130157+191511B & 13 01 55.98 & +19 15 06.26 & STAR!         &  -   &  -                                                         \\
PCG130157+191511C & 13 01 57.72 & +19 15 11.27 & 23931$\pm$ 22 & em   & [OII],H$\gamma$,H$\beta$,[OIII],[OI],H$\alpha$,[NII],[SII] \\ 
PCG130157+191511D & 13 01 58.24 & +19 15 20.38 & 23721$\pm$ 43 & em   & [OII],H$\gamma$,H$\beta$,[OIII],[OI],H$\alpha$,[NII],[SII] \\ \hline
PCG130926+155358A & 13 09 26.89 & +15 54 19.48 & 44817$\pm$ 20 & abs  &  -                                                         \\ 
PCG130926+155358B & 13 09 26.26 & +15 54 07.67 & STAR!         &  -   &  -                                                         \\ 
PCG130926+155358C & 13 09 26.44 & +15 54 12.20 & 44406$\pm$ 50 & mix  & H$\alpha$,[NII]                                            \\
PCG130926+155358D & 13 09 26.91 & +15 53 37.82 & 44896$\pm$ 20 & abs  &  -                                                         \\ \hline
PCG145239+275905A & 14 52 39.07 & +27 58 50.42 & 35907$\pm$ 35 & mix  & H$\alpha$,[NII]                                            \\ 
PCG145239+275905B & 14 52 40.31 & +27 58 48.18 & 38188$\pm$ 65 & mix  & H$\alpha$,[NII]                                            \\
PCG145239+275905C & 14 52 39.47 & +27 59 22.67 & 38019$\pm$ 54 & abs  &  -                                                         \\
PCG145239+275905D & 14 52 38.65 & +27 59 01.67 & 37490$\pm$ 71 & abs  &  -                                                         \\ \hline
PCG154930+275637A & 15 49 30.29 & +27 56 20.47 & 22910$\pm$ 18 & mix  & H$\alpha$,[NII]                                            \\ 
PCG154930+275637B & 15 49 29.11 & +27 56 41.79 & 36450$\pm$ 80 & abs  &  -                                                         \\
PCG154930+275637C & 15 49 31.49 & +27 56 40.99 & 28381$\pm$ 30 & mix  & H$\alpha$,[NII],[SII]                                      \\ 
PCG154930+275637D & 15 49 31.41 & +27 56 46.83 & 22893$\pm$ 20 & mix  & H$\alpha$,[NII]                                            \\ \hline
PCG161754+275834A & 16 17 55.29 & +27 58 32.01 & 37779$\pm$ 35 & abs  &  -                                                         \\ 
PCG161754+275834B & 16 17 54.17 & +27 58 13.98 & 37748$\pm$ 36 & abs  &  -                                                         \\ 
PCG161754+275834C & 16 17 54.86 & +27 58 12.61 & 38111$\pm$ 30 & abs  &  -                                                         \\ 
PCG161754+275834D & 16 17 55.72 & +27 58 41.48 & N/A           &  -   & strong moon reflection in the spectrum                     \\ \hline
PCG170458+281834A & 17 04 57.36 & +28 18 33.81 & 49131$\pm$ 56 & abs  &  -                                                         \\
PCG170458+281834B & 17 04 58.02 & +28 18 09.29 & STAR!         &      &  -                                                         \\
PCG170458+281834C & 17 04 59.77 & +28 18 42.95 & 24318$\pm$ 42 & mix  & H$\alpha$,[NII]                                            \\ 
PCG170458+281834D & 17 04 56.59 & +28 18 50.58 & 49431$\pm$ 59 & abs  &  -                                                         \\ \hline
\end{longtable}

\setcounter{figure}{1}
\begin{figure*}
\resizebox{\hsize}{!}{\includegraphics{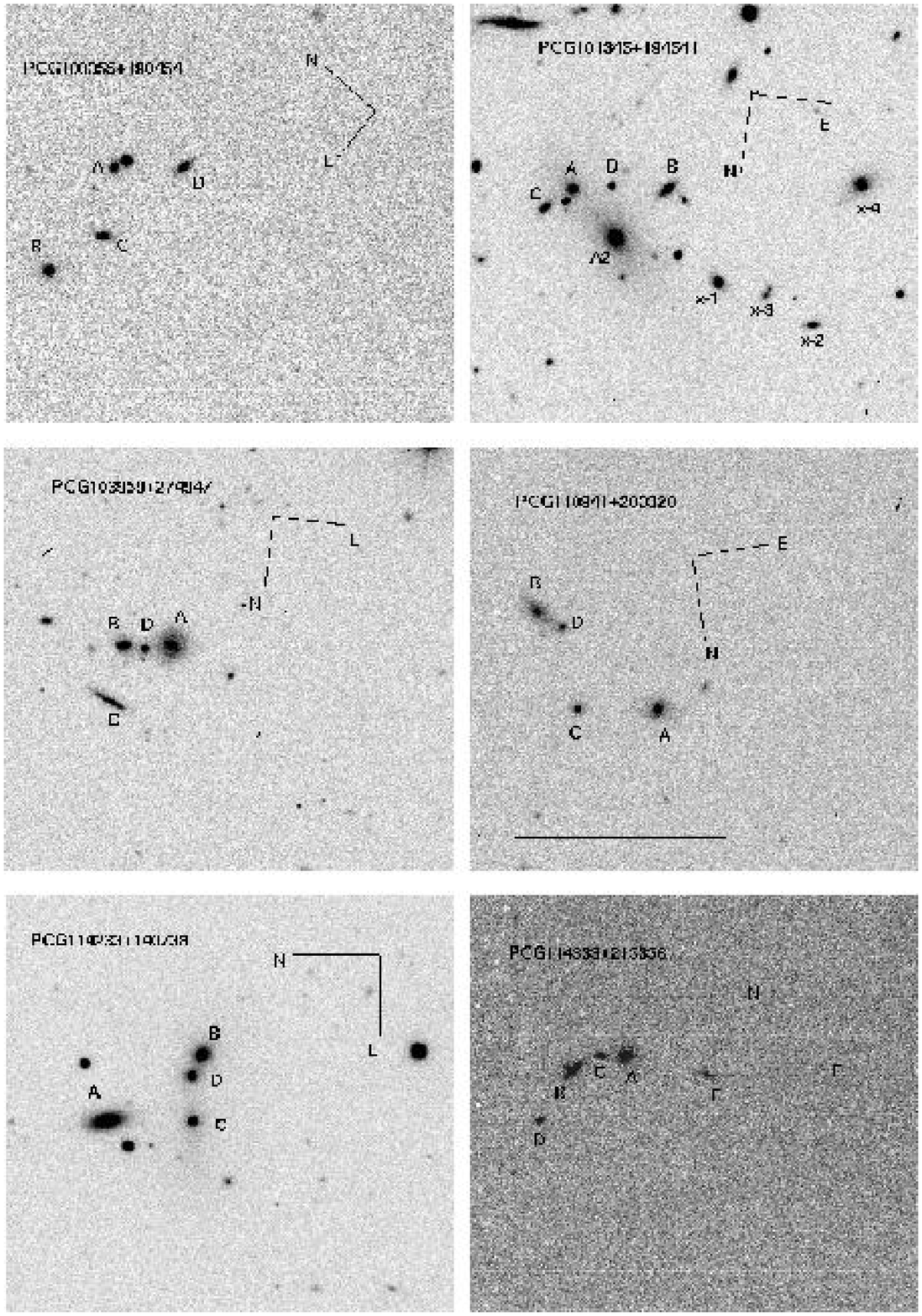}}
\caption{Acquisition images of the concordant groups. The
orientation is explicitly given for each frame. The individual
galaxies are labelled as in the DPOSS catalog. Galaxies labelled
with an x followed by a number are objects which fell in
the slit along the target galaxies and turned out to be
concordant with the DPOSS targets. The solid line
at the bottom of each image corresponds to 1$\arcmin$}
\label{Fig. 2a}
\end{figure*}

\setcounter{figure}{1}
\begin{figure*}
\resizebox{\hsize}{!}{\includegraphics{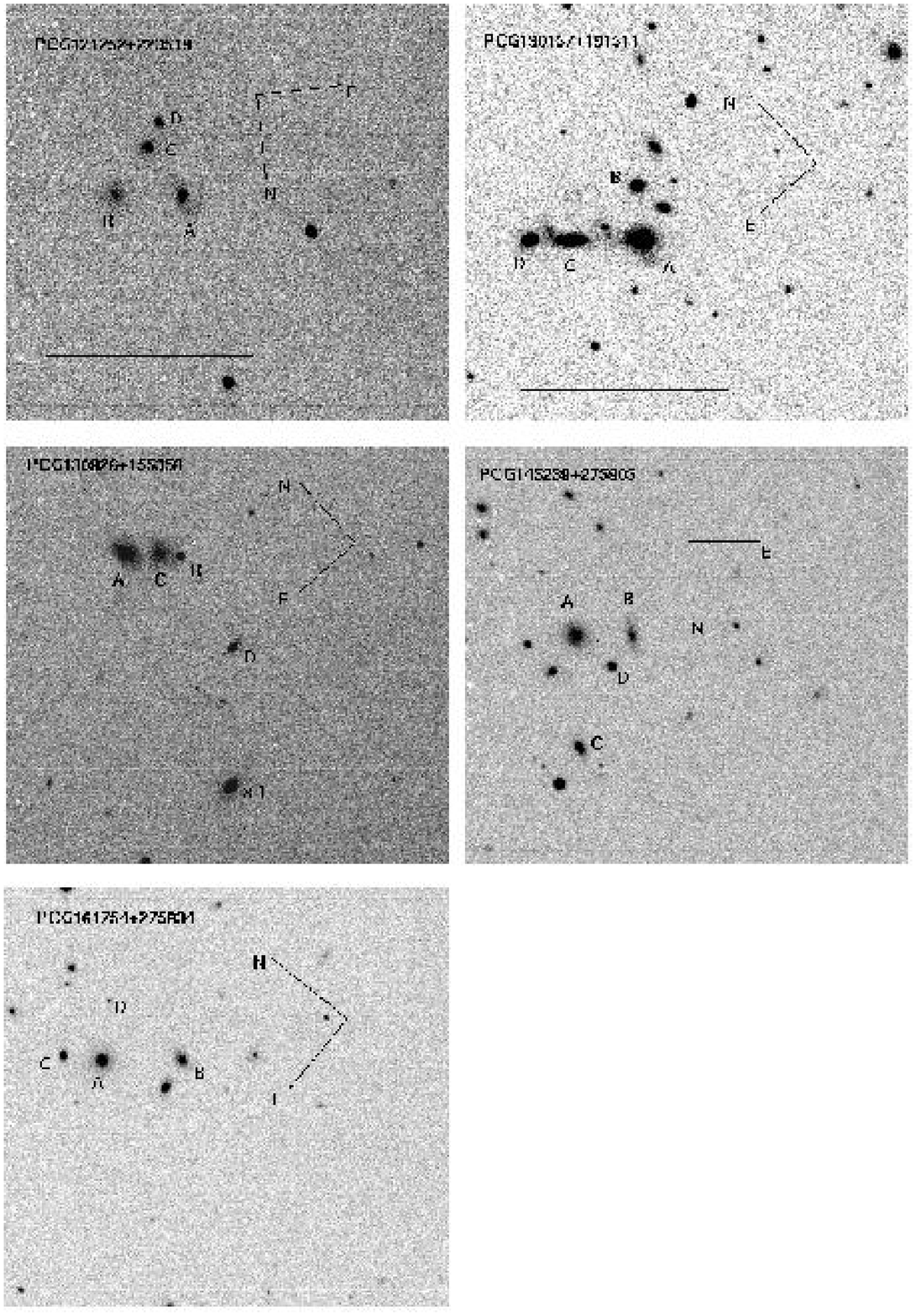}}
\caption{Fig. 2, continued}
\label{Fig. 2b}
\end{figure*}
\newpage

\setcounter{figure}{2}
\begin{figure*}
\resizebox{\hsize}{!}{\includegraphics{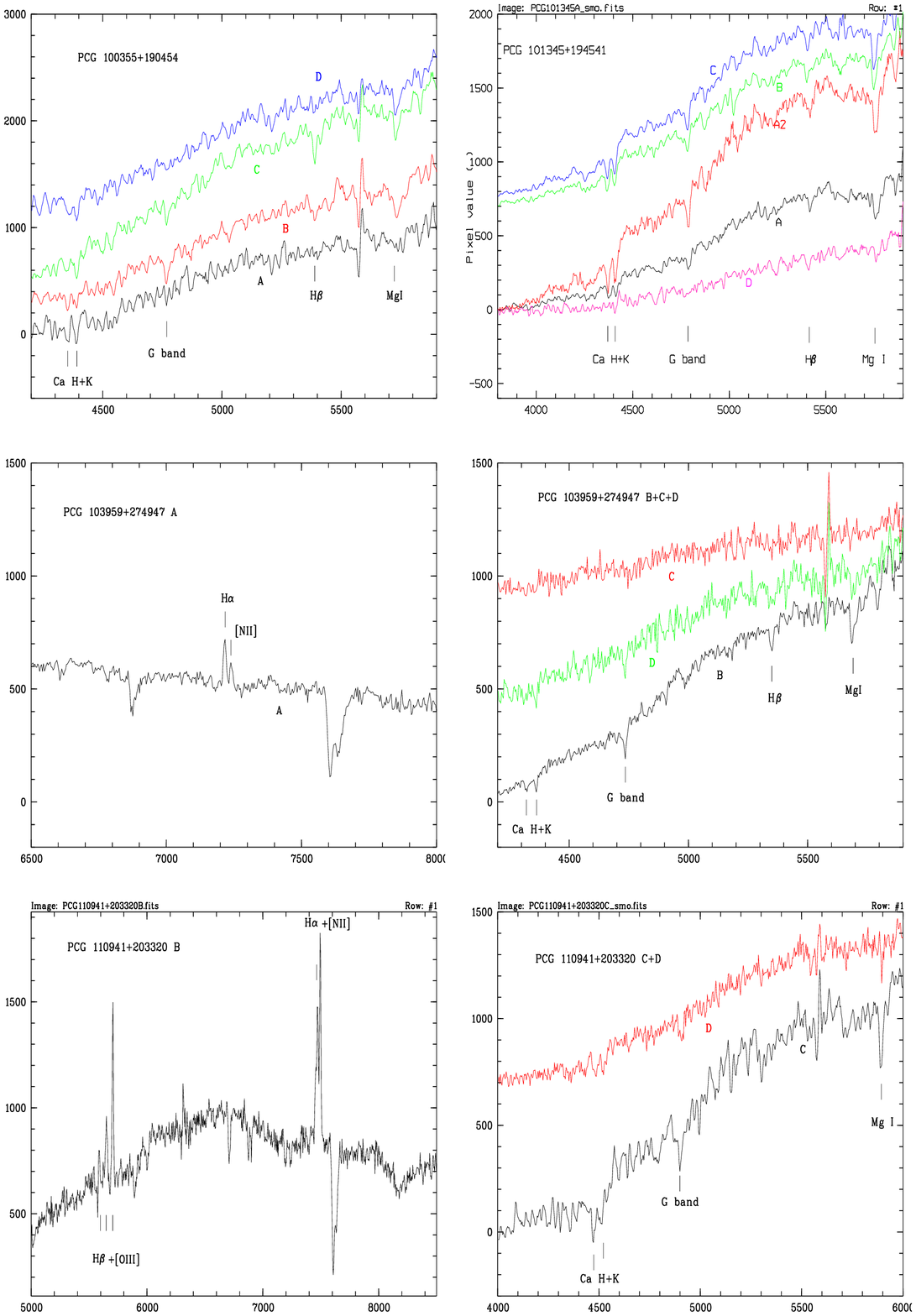}}
\caption{Spectral atlas of the concordant member galaxies for
groups of class A and B. On the x axis there is the wavelength, while on
the y axis there are the number counts (ADU). The spectra have been 
shifted an
arbitrary amount for display purpose.}
\label{Fig. 3a}
\end{figure*}

\setcounter{figure}{2}
\begin{figure*}
\resizebox{\hsize}{!}{\includegraphics{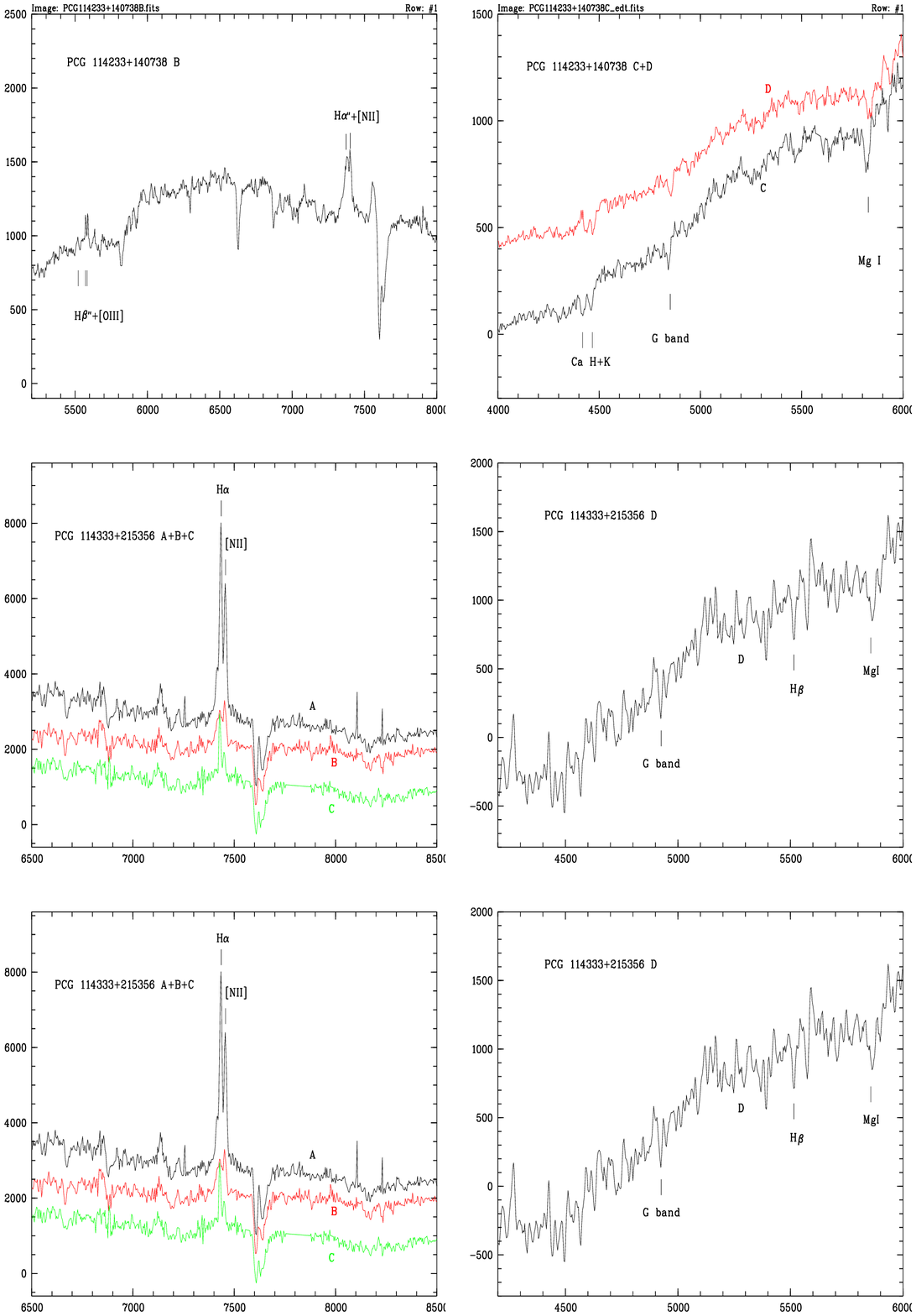}}
\caption{Fig. 3, continued}
\label{Fig. 3b}
\end{figure*}
\newpage

\setcounter{figure}{2}
\begin{figure*}
\resizebox{\hsize}{!}{\includegraphics{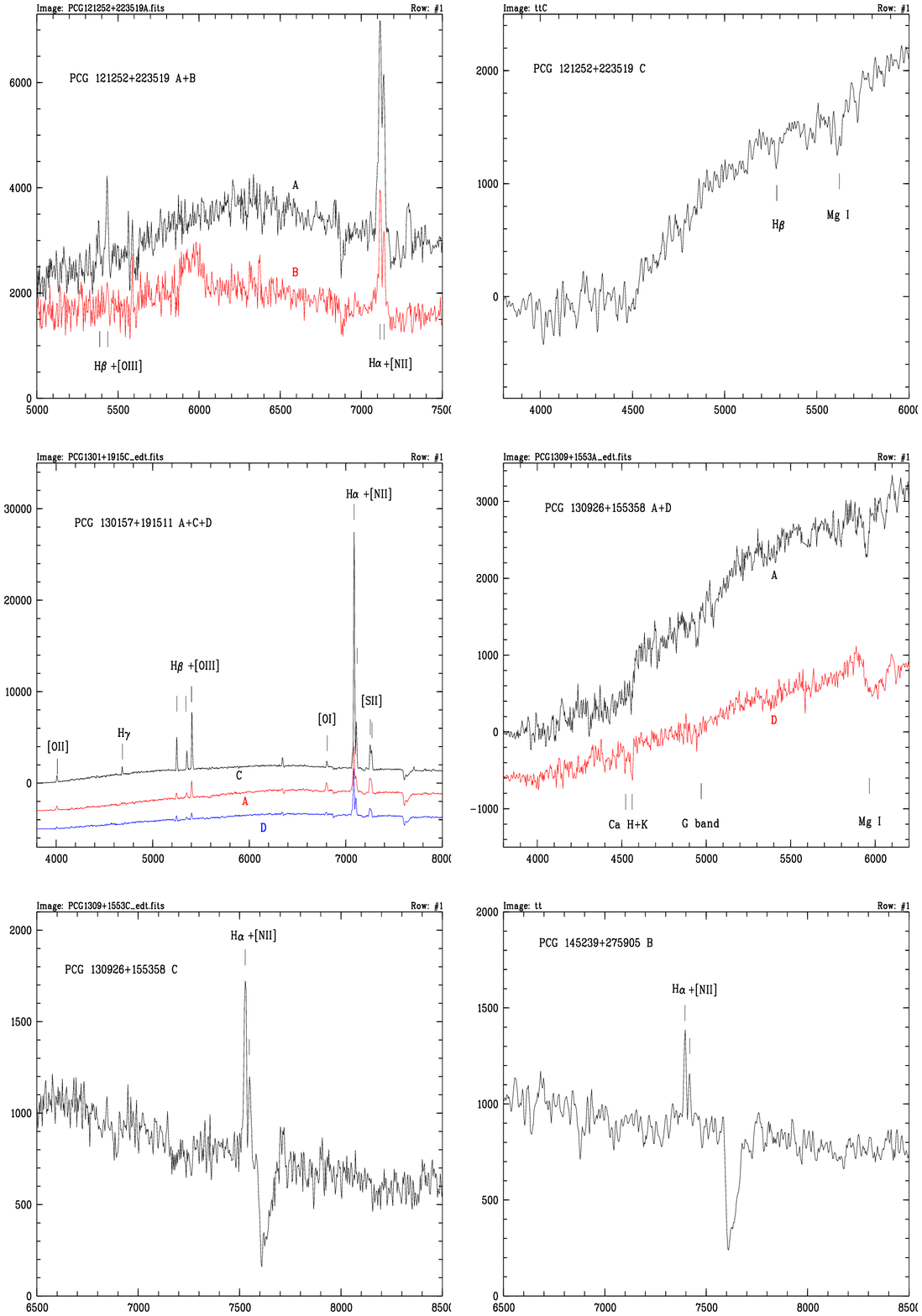}}
\caption{Fig. 3, continued}
\label{Fig. 3c}
\end{figure*}
\newpage

\setcounter{figure}{2}
\begin{figure*}
\resizebox{\hsize}{!}{\includegraphics{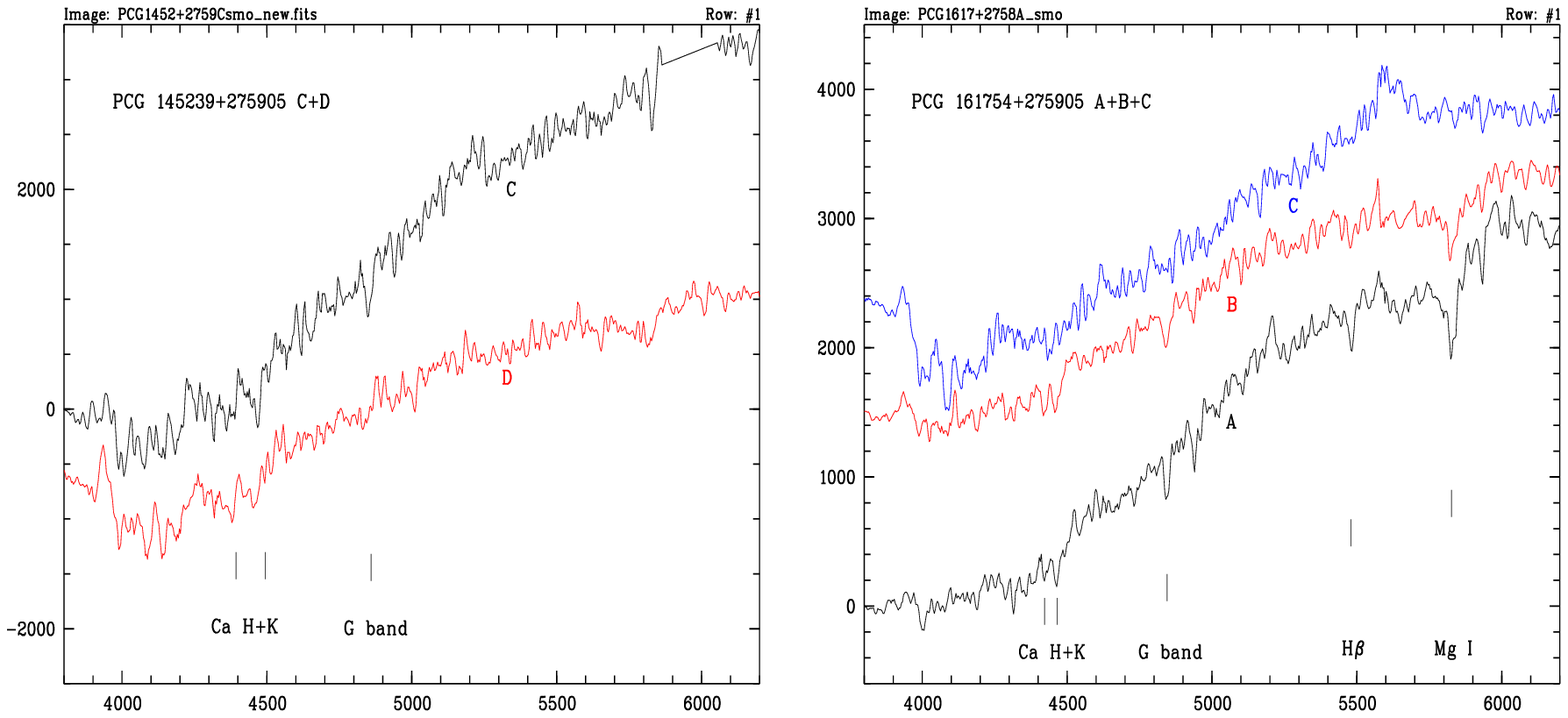}}
\caption{Fig. 3, continued}
\label{Fig. 3d}
\end{figure*}

\end{document}